\documentclass{aa}

\usepackage{graphicx}

\usepackage{txfonts}
\usepackage{xcolor}
\usepackage{hyperref}
\hypersetup{
colorlinks=true,
linkcolor=blue,
citecolor=blue,
}

\usepackage[starfontserif]{starfont}

\DeclareSymbolFont{starfontsym}{OT1}{sts}{m}{n}
\DeclareMathSymbol{\mathSun}{\mathord}{starfontsym}{115}
\DeclareMathSymbol{\mathMercury}{\mathord}{starfontsym}{102}
\DeclareMathSymbol{\mathVenus}{\mathord}{starfontsym}{103}
\DeclareMathSymbol{\mathTerra}{\mathord}{starfontsym}{76}
\DeclareMathSymbol{\mathvarTerra}{\mathord}{starfontsym}{108}
\DeclareMathSymbol{\mathMoon}{\mathord}{starfontsym}{100}
\DeclareMathSymbol{\mathvarMoon}{\mathord}{starfontsym}{97}
\DeclareMathSymbol{\mathMars}{\mathord}{starfontsym}{104}
\DeclareMathSymbol{\mathJupiter}{\mathord}{starfontsym}{106}
\DeclareMathSymbol{\mathSaturn}{\mathord}{starfontsym}{83}
\DeclareMathSymbol{\mathUranus}{\mathord}{starfontsym}{70}
\DeclareMathSymbol{\mathvarUranus}{\mathord}{starfontsym}{65}
\DeclareMathSymbol{\mathNeptune}{\mathord}{starfontsym}{71}
\DeclareMathSymbol{\mathPluto}{\mathord}{starfontsym}{74}
\DeclareMathSymbol{\mathvarPluto}{\mathord}{starfontsym}{72}

\newcommand{\kepler}{\textit{Kepler}}
\newcommand{\Nstar}{23}

\begin{document}

   \title{Seismic detection of core magnetic fields in red giants using the gravity offset}

   \author{M. Villate
          \inst{1}
          \and
          S. Deheuvels
          \inst{1,2}
          \and
          J. Ballot
          \inst{1}
          }

   \institute{%\inst{1} 
   IRAP, Université de Toulouse, CNRS, CNES, UPS, 14 avenue Edouard Belin, 31400 Toulouse, France\\
              \email{matisse.villate@irap.omp.eu, sebastien.deheuvels@irap.omp.eu, jerome.ballot@irap.omp.eu}  
    \and %\inst{2}
    Institut Universitaire de France (IUF)
    }

   \date{Received 18 December 2025, accepted 11 February 2026}

  \abstract
   {Magnetic fields are known to efficiently redistribute angular momentum in stars. They have been recently measured in the cores of red giant stars using asteroseismology. 
   It was shown that core magnetic fields, if unaccounted for, can bias the measurements of gravity-mode asymptotic properties, in particular the gravity offset $\epsilon_g$, which is otherwise well characterised for red giants. } 
   {Our aim is to exploit this bias as a way to detect magnetic fields in the cores of red giants within the \kepler\ catalogue. 
   We wish to increase the number of magnetic field detections in red giants, but also to establish a method that could be widely applicable to all red giants.}
   {We selected 218 \kepler\ red giants showing abnormal measured values of $\epsilon_{\rm g}$. We used robust statistical criteria based on the expected lifetime of mixed modes to identify significant modes.
   We then adjusted an asymptotic expression for mixed modes
   to the observed frequencies, taking magnetic field and rotation into account, using Bayesian inference. We then assessed the probability of magnetic field detection, and measured the magnetic field intensity using  stellar models for the favourable cases.}
   {We found \Nstar{} new magnetic red giant stars with fields ranging from 34 to 260~kG. For these stars, we measured values for $\epsilon_g$ now in agreement with the expected value. We also placed constraints on the magnetic field topology for seven stars. Adding these new detections to those of previous studies, we showed that the mass distribution of magnetic giants is similar to that of the complete catalogue of red giants, but different from the mass distribution of red giants with suppressed dipole modes. We also found that the core rotation of magnetic red giants follows a similar distribution as red giants in general. This could either mean that the detected fields do not have a predominant impact on the redistribution of angular momentum, or that other red giants also harbour an internal field that is currently non detectable or has not yet been detected.}
   {}


   \maketitle

\nolinenumbers

\section{Introduction}

The problem of the transport of angular momentum is one of the major obstacles of modern stellar physics. Helioseismic measurements of the internal rotation of the Sun \citep{schou_helioseismic_1998,chaplin_rotation_1999}, and later spatial asteroseismology with the CoRoT \citep{baglin_corot_2006} and \emph{Kepler} \citep{borucki_kepler_2010} missions led to a revolution in the field.
Internal rotation rates were measured for main sequence stars \citep{kurtz_asteroseismic_2014,benomar_nearly_2015,li_gravitymode_2020}, red giant stars \citep{beck_fast_2012,mosser_spin_2012,deheuvels_seismic_2014,triana_internal_2017,gehan_core_2018,kuszlewicz_Mixedmode_2023}, and white dwarves \citep{hermes_white_2017}. These results, when compared to the predictions of current stellar models for main sequence stars \citep{ouazzani_Doradus_2019} and red giants \citep{marques_seismic_2013,eggenberger_angular_2012,ceillier_understanding_2013} show that core rotation rates are largely overestimated by the models. This demonstrates that additional physical mechanisms are needed to properly explain and understand the evolution of rotation in stars. 
It was suggested that angular momentum could be transported by internal gravity waves \citep{talon_hydrodynamical_2005} or by magnetic fields \citep{eggenberger_stellar_2005} in stellar interiors.

Magnetic fields have been shown to be promising candidates for the efficient redistribution of angular momentum \citep[e.g.][]{cantiello_angular_2014,jouve_threedimensional_2015,fuller_slowing_2019,meduri_angular_2024}. They have been observed at the surface of stars for a long time \citep{landstreet_magnetic_1992,donati_magnetic_2009}, where they are generally linked to dynamo effects in convective envelopes \citep{donati_magnetic_2009}. Although magnetic fields are expected to be found in stellar interiors, detecting such fields was not possible until recently. Magnetic fields were measured for the very first time in the cores of red giant stars \citep{li_magnetic_2022,li_internal_2023,deheuvels_strong_2023,hatt_asteroseismic_2024} through their effects on the frequencies of so-called mixed oscillation modes. In red giants, non-radial modes are mixed modes, which means that they behave as pressure ($p$) modes in the envelope and as gravity ($g$) modes in the core, offering an opportunity to probe the deepest layers of the stars.
Weak magnetic fields have been very recently discovered in the core of a slowly rotating $\gamma$ Dor star \citep{takata_asteroseismic_2025}, by studying asymmetric dipolar $g$-mode multiplets and linking their existence to predominantly toroidal fields. Another study associated the nine components of a quadrupolar ($l=2$) low frequency multiplet of a $\beta$~Cep pulsator to the presence of a magnetic field in the core, and measured its strength \citep{vandersnickt_asteroseismic_2025}.

The effects of magnetic fields on oscillation modes have been studied and explored prior to their discovery. Magnetic fields, if weak enough, can be interpreted as small perturbations of the oscillation equations \citep{gough_effect_1990}. The induced perturbations on mode frequencies were computed in the specific case of a dipolar field aligned with the rotation axis for pure $g$-modes \citep{hasan_probing_2005}, and red giant mixed modes \citep{gomes_core_2020,mathis_probing_2021,bugnet_magnetic_2021}. The case of inclined dipolar fields was addressed by \cite{loi_Topology_2021}. The generalized case was addressed in \cite{li_magnetic_2022}, where the only assumption made is that the toroidal component of the field does not completely dominate over the radial component. These studies showed that weak magnetic fields cause a departure from the asymptotic expressions of $g$-mode frequencies in the form of a frequency shift that strongly depends on frequency. This shift also creates asymmetries in the rotational multiplets of non-radial mixed modes, depending on the geometry of the field. \cite{li_magnetic_2022,li_internal_2023} exploited this property of asymmetry to make the first detections of magnetic fields in red giants using dipolar ($l=1$ mixed modes). The authors measured the intensities of the fields in the core and placed constraints on their topology. \cite{deheuvels_strong_2023} detected strong, near-critical magnetic fields through their impact on the regularity of the $g$-mode pattern. \cite{hatt_asteroseismic_2024} studied low-luminosity, high signal-to-noise ratio red giants to find magnetic field signatures in mixed mode triplets. Today, we count 48 direct detections of magnetic fields in red giant stars. Other studies proposed magnetism as an explanation for the existence of a category of red giants with ``suppressed'' dipolar modes, that is, dipolar modes with unexpectedly low amplitudes (\citealt{mosser12}, \citealt{garcia14}). It was suggested that these stars harbour core magnetic fields that exceed the critical field above which magneto-gravity waves cannot propagate \citep{fuller_asteroseismology_2015,stello_prevalence_2016,cantiello_asteroseismic_2016}. 
However the interpretation on the nature and origin of suppressed modes is under debate \citep{mosser_dipole_2017}.
Super-critical fields were also invoked to account for oscillation mode suppression in a B-type main-sequence pulsator (\citealt{lecoanet_asteroseismic_2022}).

\cite{li_magnetic_2022} suggested that magnetic fields, if unaccounted for, can bias the measurement of $g$-mode parameters such as the asymptotic period spacing $\Delta\Pi_1$ and the gravity offset $\epsilon_g$. The latter parameter is linked to the properties near the turning points of the $g$-mode cavity \citep{pincon_evolution_2019}. The gravity offset was shown to have a well-determined value for all red giant stars, at $\epsilon_g = 0.28\pm0.08$ \citep{mosser_period_2018}, such that a deviation from this value is a priori not expected. The largest catalogue of seismic measurements of red giants to-date was made by \cite{li_asteroseismic_2024}. They measured the rotation and other seismic parameters (including $\epsilon_g$) for 2\,495 stars, but without taking magnetic effects into account. This makes for an interesting opportunity to test the effect magnetic fields have on the measure of $\epsilon_g$ according to the suggestion of \cite{li_magnetic_2022}.

In this study, we investigate the use of the gravity offset as an indicator for magnetic fields using the catalogue of \cite{li_asteroseismic_2024}. In Sect. \ref{Sample selection} we explain the effect the magnetic field has on the measurement of $\epsilon_g$. In Sect. \ref{analysis} we present the data used in this study, and the steps taken to extract mixed mode frequencies from the oscillation spectra. Then, in Sect. \ref{measurement} we present the asymptotic model we used to describe mode frequencies, accounting for both rotation and magnetic fields. This model is then adjusted to the data using Bayesian inference. For our magnetic detections, we measure the intensity of the field. Our results are compared to previous studies in Sect. \ref{comparison}. Lastly, in Sect. \ref{discussion} we explore the distributions of different seismic parameters for the total sample of magnetic detections.

\section{Gravity offset $\epsilon_g$ as a probe of core magnetic fields}
\label{Sample selection}

Magnetic fields have proven hard to detect in the preceding studies. Out of around 2\,000 targets, only 48 have confirmed detections of magnetic field. Increasing the number of detections is vital to the study of the origin and prevalence of magnetic fields. Past studies have established selection criteria to increase the chance to detect magnetic fields, such as looking for triplet asymmetries (\citealt{li_internal_2023}), or searching for irregularities in the measurements of the asymptotic period spacing $\Delta\Pi_1$ in \cite{deheuvels_strong_2023} for near-critical fields.
 
\begin{figure}
\centering
\includegraphics[width=\hsize]{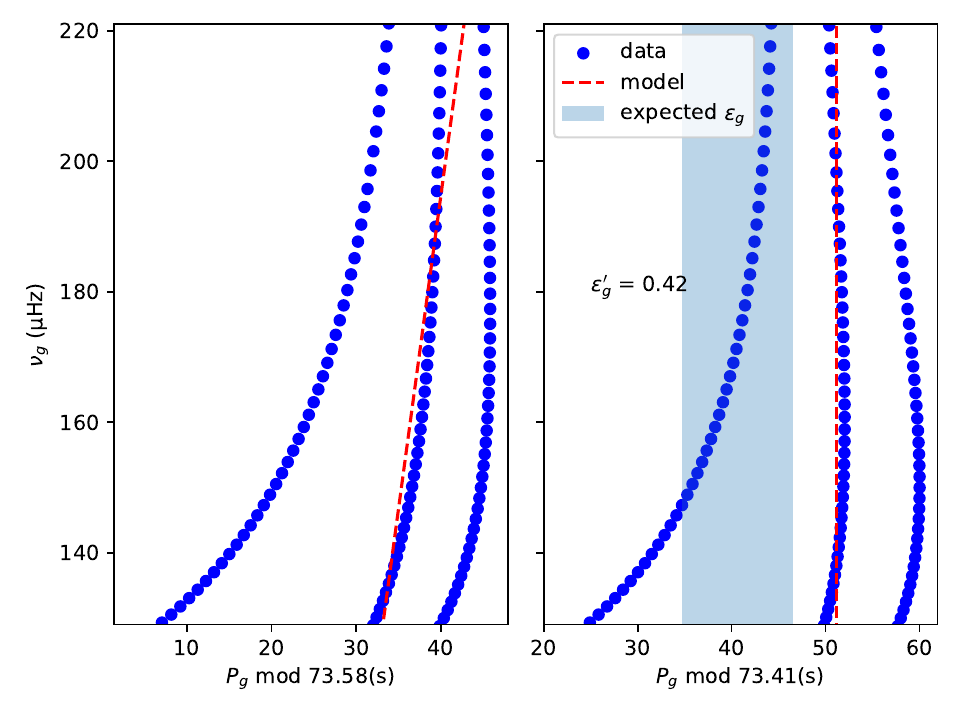}
  \caption{Échelle diagrams of mock $g$-mode spectra, perturbed by rotation and magnetic field. {\bf Left:} Spectrum fitted with an asymptotic expression accounting for magnetic and rotational perturbations, folded by the true $\Delta\Pi_1$. {\bf Right:} Spectrum fitted with an asymptotic expression for rotation only, folded by the measured $\Delta\Pi_1'$. The red lines represent the slope of the $m=0$ component, which is purely vertical on the right. The blue area represents the area where $m=0$ modes are expected to be found using $\epsilon_g = 0.28\pm0.08$}
          
     \label{shift}
\end{figure}

As presented earlier, \cite{li_magnetic_2022} suggested that weak magnetic fields, if not accounted for, can bias the measurement of g-mode parameters such as $\Delta\Pi_1$ and $\epsilon_g$. To illustrate this, we generated a set of g-mode periods using the asymptotic expression $P_g = \Delta\Pi_1(n_g+1/2+\epsilon_g)$. We then added the effects of rotation and a moderate magnetic field following a perturbative approach (this step is explained in more detail in Sect. \ref{Model}). The resulting mode periods are represented in the left panel of Fig.~\ref{shift} in the shape of an échelle diagram folded with the true period spacing $\Delta\Pi_1$. The two main effects of magnetic fields appear clearly. Since magnetic frequency shifts strongly depend on frequency ($\sim \omega^{-3}$, see \citealt{mathis_probing_2021}, \citealt{li_magnetic_2022}), low-frequency modes are more affected, resulting in a bending of the ridges to the left at low-frequency in the échelle diagram. Secondly, the shift differs for $m=0$ and $m=\pm1$ components, which creates an asymmetry of the triplet. If magnetic fields are ignored, the $m=0$ modes are fitted as $P'_n = \Delta\Pi_1'(n_g+1/2+\epsilon_g')$, where the apparent $\Delta\Pi_1'$ is chosen so that the $m=0$ ridge is as close as possible to a straight vertical line in the échelle diagram (red dashed line in Fig. \ref{shift}). The result is shown in an échelle diagram folded with the apparent period spacing $\Delta\Pi_1'$ in the right panel of Fig. \ref{shift}. For a weak field, the value of $\Delta\Pi_1'$ differs from $\Delta\Pi_1$ by only a few tenths of s, which is not enough to make these measurements abnormal. On the contrary, the value of $\epsilon_g$ is much more affected, and $\epsilon_g'$ can significantly differ from the expected value of $0.28 \pm 0.08$. As seen in the right panel of Fig. \ref{shift}, the blue area represents the zone in which we would expect to find most modes considering the empirical value of $\epsilon_g$. This is not the case here where the measured $\epsilon_g'$ is subject to significant distortions (it corresponds to $\epsilon_g' = 0.42$ in the example shown in Fig. \ref{shift}).

\begin{figure}
\centering
\includegraphics[width=\hsize]{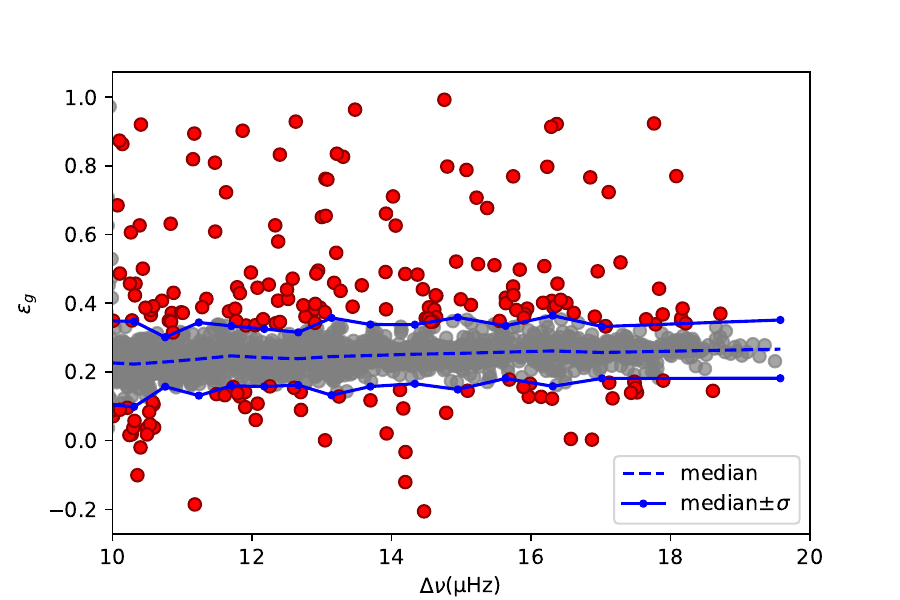}
  \caption{Relationship between $\epsilon_g$ and $\Delta\nu$ in the \cite{li_asteroseismic_2024} catalogue. Red dots represent stars selected for this study, in contrast to the grayed out dots. Full blue lines represent the delimitation of the bulk distribution at $\pm \sigma$ and the dashed blue line is made from the median of every 118 stars slices.
          }
     \label{sample}
\end{figure}

\cite{li_asteroseismic_2024} measured mixed mode parameters for 2\,006 red giants with great precision. The authors also measured core and envelope rotation rates for red giants showing at least two components per dipole triplet. However, these measurements did not take magnetic fields into account. In Fig. \ref{sample}, we show the measured values of $\epsilon_g$ as a function of the large separation $\Delta\nu$ of $p$-modes. Clearly, most stars are found in the expected range for $\epsilon_g$, but a smaller population of red giants have $\epsilon_g$ that clearly deviate from the empirical value (these stars are highlighted in red in Fig. \ref{sample}).
Following the behaviour explained above, these stars are interesting candidates for hosting a magnetic field.

To select these candidates, we first calculated the distribution of $\epsilon_g$ in slices of $\Delta\nu$, each slice containing 118 stars. The standard deviation and median were then calculated for each slice, and used to delimit the bulk of the full distribution. The stars that lie outside of $\pm 1\sigma$ of the delimitation were taken as part of the sample. Also, in this study, we chose to focus on low-luminosity red giants, using as a constraint $\Delta\nu>10\,\mu\mathrm{Hz}$. Indeed, for higher-luminosity giants, the oscillation spectra become more complex, the rotational splitting being generally larger than the frequency separation between mixed modes. Higher-luminosity stars will be treated in a later study. In total, we selected 218 low-luminosity red giants with abnormal values of $\epsilon_g$ for analysis.

\section{Analysis of oscillation spectra}
\label{analysis}
\subsection{Seismic data and pre-processing}
\label{seismic}

In this study, we used the full \kepler\ \citep{borucki_kepler_2010} dataset, corresponding to nearly 4-year long high-precision photometry time series. The oscillation spectra were downloaded from the Mikulski Archive for Space Telescopes (MAST). We used the calibrated KEPSEISMIC\footnote{\url{http://dx.doi.org/10.17909/t9-mrpw-gc07}} data, high-pass filtered at 20~days \citep{garcia_Preparation_2011,garcia_impact_2014,pires_gap_2015}. We then fitted to the observed spectra a model of the background in order to determine the signal-to-noise ratio (SNR) of the Power Spectral Density (PSD) of each star. The background was fitted using two Harvey laws for the granulation, a white noise component for photon noise, and a Gaussian function corresponding to the pressure mode bump. 
The fit was done with a Maximum Likelihood estimation (MLE),
using \texttt{apollinaire} \citep{2022A&A...663A.118B}. Initial guesses for this estimation were chosen based on the results of previous fits for similar stars.

\subsection{Extraction of mode frequencies}

The signature of a magnetic field is found either by detecting asymmetry in the dipole triplet (only if three components are visible) or by measuring deviations from a constant g-mode period spacing. This translates into curved ridges in the stretched échelle diagram (similarly to Fig.~\ref{shift}), where stretching designates a transformation of mixed mode periods to account for the variation of period spacing \citep[see][]{mosser_period_2015}. In the latter case, the detection of a field crucially depends on modes in the low-frequency part of the observed spectrum because magnetic effects are largest at low frequency. However, these modes also have lower SNR. For this reason, special care is needed to (i) maximise the probability of detection of lower-frequency modes, and (ii) limit as much as possible the false-alarm probability, in order to avoid false detections of magnetic fields. Many seismic studies assume an empirical detection threshold of eight times the background noise level \citep[e.g.][]{mosser_period_2015}. In this work, we developed a more robust detection criterion.

We first elaborated a basic, conservative SNR threshold over which the signal is considered as a detection. To do so, we considered each bin in the oscillation spectrum as a realisation of a random variable $X$ following a $\chi^2$ distribution with two degrees of freedom \citep[e.g.][]{1998A&AS..132..107A}. Here two possible outcomes exist: either the realisation of $X$ is due to noise (hypothesis $H_0$), or it is signal (hypothesis $H_1$). We want to define a threshold $x_p$ in the PSD over which peaks are considered to be due to signal. The false-alarm probability, known as the $p$-value, corresponds to the probability of having at least one peak due to noise that exceeds $x_p$ among the $N$ considered frequency bins, that is, $P(\;\exists i \in [\![1,N]\!] \;;\; X_i > x_p\;|\;H_0) = p $. For a given $p$-value, assuming that all the bins are independent from one another (this is a good approximation because the duty cycle of \kepler\ is above 90\%), the threshold $x_p$ can be expressed as \citep[e.g.][]{deheuvels_thesis}:
\begin{equation}
    x_p = - \ln \left[1-(1-p)^{1/N} \right]
    \label{eq:threshold}
\end{equation}
We consider modes in the frequency range $\nu_\mathrm{max}\pm5\Delta\nu$. For 4-yr-long \textit{Kepler} datasets (with a frequency resolution of $\sim 7\,\mathrm{nHz}$), this selects $N\in [14\,000;25\,000]$ bins for each star. For a $p$-value of 5\%, we obtain $x_p \approx 12$. This threshold guarantees a 95\% chance of having no peak due to noise exceeding $x_p$ in the considered frequency range. 

For stochastically excited modes, the mode lifetime will generally be shorter than the duration of the observations, so that these modes are spread over several bins of the spectrum. We can concentrate the power of a mode by smoothing the spectrum, redistributing power into fewer bins, to maximise detection probability \citep{appourchaux_detecting_2004}. The width of the smoothing window, thereafter ``boxcar width'', is usually taken as a multiple of the mode width. Two courses of action are possible when we smooth the spectrum. Either re-binning the spectrum over the $N$ bins of the boxcar, or keeping every bin, in which case we loose the independence between bins and the threshold cannot be computed analytically. Here, we chose the latter because the first option yields a significantly worse detection probability. We thus used a Monte Carlo approach to infer the threshold in SNR corresponding to a false alarm probability of $5\%$ for each boxcar width. We generated series of pure noise (random 2-dof $\chi^2$) with the same number of bins as the real spectrum, and we applied a smoothing over the chosen boxcar. This procedure was repeated over 1000 iterations. Using this, the threshold is given by the height in SNR that only $5\%$ of the realisations exceed.

\begin{figure}
\centering
\includegraphics[width=\hsize]{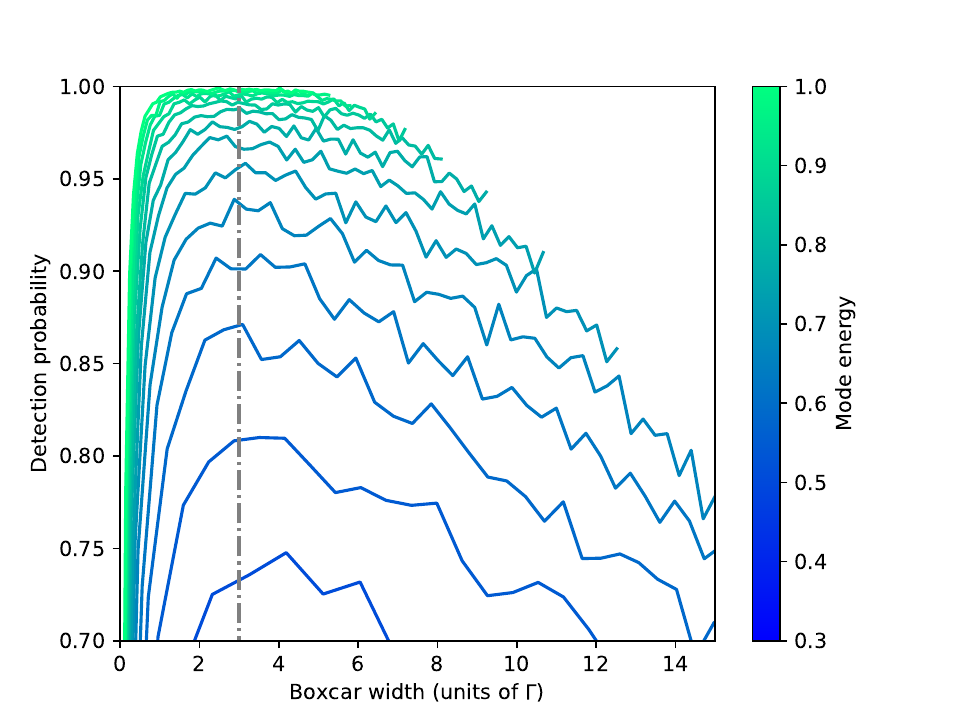}
  \caption{Probability of detection of a mode versus the boxcar width, in units of the mode width. The colour of the lines indicates the mode energy, defined as $E \propto \Gamma h$ where $\Gamma$ is the mode width and $h$ is mode height. The vertical dash-dotted line corresponds to a boxcar width of $3\,\Gamma$.
          }
     \label{boxcar}
\end{figure}

We then searched for the boxcar width that maximizes our chances of detecting the mode. For this purpose, we generated synthetic spectra containing a single mode, to which we applied noise following a 2-dof $\chi^2$ distribution. This procedure was repeated over $10\,000$ iterations.
For different mode energies (proportional to the product of mode height and width), we measure the detection probability for several boxcar widths. Using this, we were able to determine that the probability of detection is at its peak when the boxcar width is about three times the mode width (see Fig.~\ref{boxcar}). We used this boxcar size in our subsequent analysis.

\begin{figure}
\centering
\includegraphics[width=\hsize]{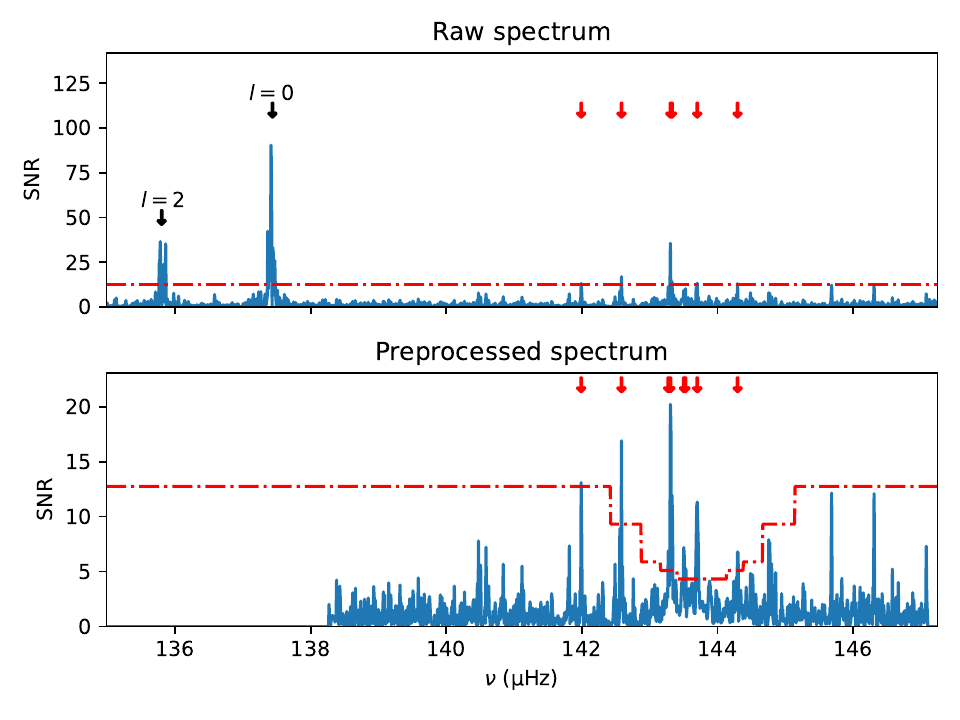}
  \caption{Spectrum of KIC9999911. {\bf Top:} Unprocessed spectrum. {\bf Bottom:} Preprocessed spectrum, with variable width smoothing and radial and quadrupolar modes cut out. Red arrows indicate the position of detected modes using the red dashed-dot threshold.
          }
     \label{smooth}
\end{figure}

In our case, the lifetimes of mixed modes vary depending on the ratio of mode inertia stored in either the acoustic cavity of the buoyancy cavity. Modes that are more $g$-like have longer lifetimes and have narrow mode profiles in the PSD, with a width close to the frequency resolution, while $p$-dominated modes span several bins of the spectrum. The width of a mixed mode can be expressed as \citep{mosser_period_2015}:
\begin{equation}
    \Gamma_{l=1} = \Gamma_{l=0}(1-\zeta) \,,
\end{equation}
where $\zeta$ is the ratio of mode inertia stored in the buoyancy cavity to the total mode inertia, which varies with frequency and can be computed using the measured frequencies of mixed modes \citep{goupil_Seismic_2013}. Here, we compute the expected trapping of the modes using parameter $\zeta$, determined using the parameters in \citet{li_asteroseismic_2024}. Using this we smooth the PSD with a boxcar width of $3\,\Gamma_{l=1}$ . We then use the first Monte Carlo scheme described earlier to compute the threshold at each boxcar width, with respect to its corresponding frequency. This results in a threshold that varies with the smoothing of the spectrum. After removing the radial and quadrupolar modes from the spectra, we show an example of this smoothing and threshold in Fig.~\ref{smooth}, which presents a portion of the spectrum of KIC9999911. Clearly, modes are more smoothed in $p$-mode regions ($\sim 142.5$ to $145 \,\mu\mathrm{Hz}$) than in $g$-mode regions (the rest of the spectrum), where it is not smoothed. We can also notice how the variable threshold allows the detection of weaker $p$-mode components (we detect one component at $\sim 143.5 \,\mu\mathrm{Hz}$ that was not detected with a flat threshold).

Using this new threshold, we were able to obtain a list of reliable $l=1$ and $l=3$ mode frequencies guesses from the spectra. In practice, several peaks sometimes belong to the same mode and obtaining one frequency guess per mode requires an additional step, which is detailed in Appendix \ref{extraction}. 

The last step consists in extracting the mode frequencies from the PSD using the guesses obtained above. For this purpose, we modeled each mode as a Lorentzian profile with free central frequency, height, and width. The background was fixed to the estimate that was made in Sect.~\ref{seismic}, using the contributions from granulation and the photon noise. The model of spectrum was fit to the PSD for each overtone with detected modes using maximum likelihood estimation (MLE) similarly to what was done, e.g., in \cite{deheuvels_seismic_2020}. The errors were then estimated by computing the covariance matrix as the inverse of the hessian matrix.

\section{Measurement of internal magnetic fields}
\label{measurement}
\subsection{Asymptotic expression of mixed mode frequencies with rotation and magnetic fields}
\label{Model}
We used the asymptotic expression for dipolar mixed mode frequencies in the oscillation spectrum, assuming the validity of the WKB approximation. \cite{shibahashi_modal_1979} and \citet{unno_nonradial_1989} have shown that the mixed mode frequencies $\nu$ can be approximated by solving:
\begin{equation}
     \tan{\theta_{p,m}(\nu)} = q \tan{\theta_{g,m}(\nu)}\,, \label{mix}
\end{equation}

where $q$ is the coupling factor, $\theta_{p,m}(\nu)$ and $\theta_{g,m}(\nu)$ are the $p$ and $g$ mode phases respectively, and the subscript $m$ refers to the dependence in the azimuthal order. Following \citet{mosser_period_2015}, we express $\theta_{p,m}$ and $\theta_{g,m}$ as
\begin{alignat}{2}
    \theta_{p,m}(\nu) &= \pi\frac{\nu-\nu_{p,m}}{\Delta\nu(n_p)}\quad, & \quad \theta_{g,m} &= \frac{\pi}{\Delta P_g(n_g)}\left(\frac{1}{\nu}-P_{g,m}\right)-\frac{\pi}{2}\,,\label{phases} 
\end{alignat}
with $\Delta\nu(n_p) = \nu_{p,m}(n_p+1)-\nu_{p,m}(n_p)$ the local frequency large separation at radial order $n_p$ and $\Delta P_g = P_{g,m}(n_g+1) - P_{g,m}(n_g)$ the local period spacing at radial order $n_g$. Frequencies $\nu_{p,m}$ are linked to the pure dipolar unperturbed $p$ mode frequencies $\nu_p^{(0)}$, which can be expressed using an asymptotic expression of $p$-modes as \citet{mosser_period_2015}:
\begin{equation}
    \nu_p^{(0)} = \left[ n_p + \frac{1}{2} + \epsilon_p + \frac{\alpha}{2}(n_p-n_\mathrm{max})^2 \right]\Delta\nu - \delta\nu_{01} \,,\label{nup2}
\end{equation}
with $\epsilon_p$ the $p$ mode offset, and $\alpha$ and $\delta\nu_{01}$ are parameters introduced by the second-order asymptotic expression. The parameter $\delta\nu_{01}$ characterises the small separation between $l=0$ and $1$ modes, and $n_\mathrm{max} = \nu_\mathrm{max}/\Delta\nu$. The periods $P_{g,m}$ are linked to the pure dipolar unperturbed $g$ mode periods $P_g^{(0)}$: 
\begin{equation}
    P_g^{(0)} = \Delta\Pi_1 (n_g+1/2+\epsilon_g)\,\label{Pg}.
\end{equation}
Equations~\ref{phases} and \ref{Pg} appear to be slightly different from the ones written by \citet{mosser_period_2015}, but they are strictly equivalent and provide the same spectrum of mixed modes. However, with our expressions, $g$-dominated mixed mode frequencies tend to $g$-mode frequencies when $q$ tends to zero. 

The frequencies $\nu_{p,m}$ and periods $P_{g,m}$ contain the contributions of rotation and magnetic field as perturbations to the model, which we will now formulate.
Since red giants are slow rotators, rotation can indeed be treated as a first-order perturbation. In this case, dipolar modes are split into three components, forming a symmetrical triplet. The frequency shifts induced by rotation between each component can be expressed as $\nu_{R,p} = \langle\Omega\rangle_p/(2\pi)$ and $\nu_{R,g} = \langle\Omega\rangle_g/(4\pi)$ for $p$ and $g$ modes, respectively, and are related to the average rotation rate $\langle\Omega\rangle$ of their respective cavities  \citep[see][]{goupil_Seismic_2013}. 

Magnetic fields can also be treated as first-order perturbations if the field strength $B$ stays small compared to the critical field $B_c$. 
Several studies investigated special cases of magnetic fields with specific geometries \citep{gomes_core_2020,mathis_probing_2021,bugnet_magnetic_2021,loi_Topology_2021}.
The generalized case was treated in \cite{li_magnetic_2022}, where the only assumption is that the toroidal component of the field does not completely dominate over the radial component. 
These expressions are valid as long as $B_\phi/B_r \ll N/\Omega \sim 100$ in red giants, where $N$ is the Brunt-Väisälä frequency and $\Omega$ is the rotation of the star. In the most general case, when the effects of non-axisymmetry of the field are strong enough, each component of the dipolar mode may be divided in 3 components, with a total of 9 components. \cite{li_magnetic_2022} showed that, when the rotational splitting is smaller than the magnetic shift such that $b=\nu_B/\nu_R < 1$ , non axisymmetric effects become negligible, and frequencies behave as if the field were axisymmetric. In this study, we make the hypothesis that this property is always verified and we assess the validity of that assumption in Sect. \ref{nonaxi}. 

In our context, the frequencies of $p$ modes $\nu_{p,m}$ and the periods of $g$ modes $P_{g,m}$ perturbed by rotation and magnetic field can be written as:
\begin{alignat}{2}
    \nu_{p,m} &= \nu_p^{(0)} - m\nu_{R,p}  \quad\text{and}\quad P_{g,m} &=\left(\frac{1}{P_g^{(0)}}-m\nu_{R,g}+f_m(a)\nu_B\right)^{-1}\,,
    \label{pert}
\end{alignat}
where the term $f_m(a)\nu_B$ represents the magnetic perturbation. Magnetic fields were shown to affect mostly $g$-mode oscillations \citep{mathis_probing_2021}, which is why we did not consider a magnetic perturbation to $p$-mode frequencies. Following \citet{li_magnetic_2022}, we can express the magnetic perturbation term of Eq.~\ref{pert} as:
\begin{alignat}{2}
    \nu_B &= \nu_\mathrm{mag} \left(\frac{\nu_\mathrm{max}}{\nu}\right)^3\quad\text{and}\quad f_m(a) = 1-a\left(1-\frac{3}{2}|m|\right)\,,
\end{alignat}
where $a$ is the asymmetry parameter, linked to the latitudinal structure of the field \citep{li_magnetic_2022}. The value of $a$ varies between $-0.5$ and $1$, for a field concentrated at the equator and at the poles, respectively. The parameter $\nu_\mathrm{mag}$ is the average magnetic frequency shift at $\nu_\mathrm{max}$, which is linked to the intensity of the magnetic field by the expression:
\begin{equation}
    \nu_\mathrm{mag} = \frac{\mathcal{I}}{2\pi\mu_0(2\pi\nu_\mathrm{max})^3}\left<B_r^2\right>\,, \label{mag1}
\end{equation}
$\mathcal{I}$ being an integral depending on the core structure, and $\langle B_r^2\rangle$ an average of the radial component of the magnetic field squared weighted by the kernel function $K_r(r)$. It is defined as:
\begin{equation}
    \left<B_r^2\right> = \int_{r_1}^{r_2}K_r (r)\overline{B_r^2}\,\hbox{d}r,
    \label{kernel}
\end{equation}
where $\overline{B_r^2}$ is the average of the field in the horizontal direction. Also, as $K_r(r)$ peaks in the hydrogen burning shell (HBS) around the core,  $\langle B_r^2 \rangle$ has a high sensibility to the field in this region (see extended figure 1 from \citealt{li_magnetic_2022}).

In practice, the visibility of mode components impacts the measurement of the magnetic perturbation. Indeed, in the best-case scenario of a triplet, both $\nu_\mathrm{mag}$ and $a$ are measurable. But in other cases, only the product of $f_m(a)\nu_\mathrm{mag}$ is measurable. In the cases of a doublet or singlet we measure:
\begin{alignat}{2}
    \tilde{\nu}_\mathrm{mag} = &f_{\pm 1}(a) \nu_\mathrm{mag} &= (1+a/2)\nu_\mathrm{mag} \\
    \tilde{\nu}_\mathrm{mag} = &f_0(a)\nu_\mathrm{mag} &=  (1-a) \nu_\mathrm{mag} 
\end{alignat}
where the tilde denotes the observable parameter. Since $a$ is bounded between $-1/2$ and $1$, $\nu_\mathrm{mag}$ lies between $\frac{2}{3}\tilde{\nu}_\mathrm{mag}$ and $\frac{4}{3}\tilde{\nu}_\mathrm{mag}$ for doublets, and above $\frac{2}{3}\tilde{\nu}_\mathrm{mag}$ for singlets.

To obtain mixed mode frequencies, we have to solve Eq.~\ref{mix} for each azimuthal order $m$, and considering the perturbations of the rotation and magnetic field introduced by Eq. \ref{pert}. In the next subsection, we detail the process used to adjust these modeled frequencies to the observations.

\subsection{Bayesian inference}
\label{inference}

We fit the observed frequencies of radial and dipolar modes with the asymptotic model described in Sect.~\ref{Model}.
The free parameters of the model are those describing the $p$-mode spectrum ($\Delta\nu$, $\epsilon_p$, $\alpha$ and $d_{01}$), the dipolar $g$ mode spectrum ($\Delta\Pi_1$ and $\epsilon_g$), the coupling parameter $q$, the mean rotation rates in the $p$ and $g$ cavities ($\nu_{R,p}$ and $\nu_{R,g}$) and the core magnetic field ($\nu_\mathrm{mag}$ and $a$). We fix $\nu_\mathrm{max}$ at the value measured by \citet{yu_asteroseismology_2018}. These parameters are gathered in a parameter vector denoted $\boldsymbol{\theta}$. The modeled frequencies are generated using equations \ref{mix} to \ref{Pg} over a given range of $n_p$ and $n_g$. This range is tailored to cover the whole observed spectrum.
In order to estimate the parameters, we use a Bayesian inference approach. Following Bayes' theorem, the posterior distribution of the parameters $\Vec{\theta}$ given a data set $\Vec{y}$ --which consists here in a set of observed frequencies-- is
\begin{equation}
    p(\Vec{\theta}|\Vec{y}) = \frac{p_I(\Vec{\theta})p(\Vec{y}|\Vec{\theta})}{p(\Vec{y})}
\end{equation}
where $p_I(\Vec{\theta})$ is the prior distribution of $\Vec{\theta}$, $p(\Vec{y}|\Vec{\theta})$ the likelihood and $p(\Vec{y}) = \int p_I(\Vec{\theta})p(\Vec{y}|\Vec{\theta}) \hbox{d}\boldsymbol{\theta}$ is the global likelihood. The likelihood $p(\Vec{y}|\Vec{\theta})$ is expressed
\begin{equation}
    p(\Vec{y}|\Vec{\theta}) = \frac{1}{\sqrt{(2\pi)^n \det \Vec{\Sigma}}} \exp\left[-\frac{1}{2} [\Vec{y}-\Vec{\mu}(\Vec{\theta})]^T \Vec{\Sigma}^{-1} [\Vec{y}-\Vec{\mu}(\Vec{\theta})]\right],
\end{equation}
where $\Vec{y}=\left(\nu_{i,\mathrm{obs}}\right)_{i=1,n}$ is a vector containing the $n$ observed frequencies. The observed modes are assigned to the closest modeled modes. $\Vec{\mu}(\Vec{\theta})=\left(\nu_i(\Vec{\theta})\right)_{i=1,n}$ is a vector containing the frequencies modelled with a set of parameters $\Vec{\theta}$, and $\Vec{\Sigma}=\Vec{\Sigma}_{\rm obs}+\Vec{\Sigma}_{\rm mod}$ a covariance matrix composed of two terms. The first one, $\Vec{\Sigma}_{\rm obs}$, is the observational error covariance matrix. As the frequency measurements are independent, it boils down to a diagonal matrix $\Vec{\Sigma}_{\rm obs}=\mathrm{diag}\left(\sigma_{1,{\rm obs}}^2,\dots,\sigma_{n,{\rm obs}}^2\right)$, where $\sigma_{i,{\rm obs}}$ is the error of $\nu_{i,{\rm obs}}$. The second term, $\Vec{\Sigma}_{\rm mod}$, takes into account the errors of the model. 
Indeed, as the asymptotic expression presented in Sect.~\ref{Model} are derived from simplified versions of the equations of stellar oscillation, we expect deviations from these expressions. 
To take them into account, \citet{li_asteroseismic_2024} added a fixed error. For this work, we modelled $\Vec{\Sigma}_{\rm mod}$ as a diagonal matrix with terms 
\begin{equation}
    \sigma^2_{i,\mathrm{mod}} = \zeta_i \sigma_g^2 +(1-\zeta_i) \sigma_p^2
\end{equation}
on its diagonal. We introduced two free parameters, $\sigma_g$ and $\sigma_p$, which characterise the errors on $g$- and $p$-modes, respectively. The quantity $\zeta_i$ is the ratio of inertia in the $g$ cavity over the total inertia for the mode with frequency $\nu_i$. It is evaluated from the parameters $q$, $\Delta\Pi_1$ and the $p$-mode frequencies $\nu_p$ \citep[see, e.g., Eq.~4 from][]{gehan_core_2018}. The parameters $\sigma_g$ and $\sigma_p$ are estimated during the fitting process. This is a simplified model since the model errors of close frequencies are correlated and their distribution is not necessarily normal. Thus, we performed tests on realistic synthetic spectra to verify that the errors of parameters $\Vec{\theta}$ correctly take into account deviations from the asymptotic regime. Our approach appears to be quite conservative and tends to overestimate the errors.

The prior distributions $p_I(\Vec{\theta})$ capture all our \emph{a priori} knowledge (or lack thereof) on those parameters.
Compiled in Table~\ref{table:prior}, the priors were chosen in order to cover the known range of the parameter values. If this range is not or badly known, a large and uninformative prior was chosen for that parameter. Four types of prior distributions were used: uniform over the interval $[x_{\rm min},x_{\rm max}]$, ``uniform-periodic'', which is a uniform prior for which the resulting posteriors are the results of steps by the walkers modulo the period, ``uniform-normal'', which is a uniform distribution with Gaussian edges of width $\sigma_\mathrm{min}$ and $\sigma_\mathrm{max}$, and a modified Jeffrey prior, which is an uninformative, scale-invariant prior over $]x_{\rm min},x_{\rm max}]$, and smoothly transitions to a quasi-uniform distribution over $[0,x_{\rm min}]$ \citep[see detailed descriptions in, e.g.,][]{handberg_bayesian_2011}. These appellations are abridged in Table \ref{table:prior} as ``U'', ``UP'' , ``UG'', and ``J'' respectively.

For the parameters $\epsilon_p$ and $\epsilon_g$, which have a periodic behaviour (see Sect.~\ref{Sample selection}), we used a uniform-periodic distribution. For $\Delta\nu$, we used a uniform prior around the value measured by \cite{li_asteroseismic_2024} (noted $\Delta\nu_0$), with a width of 6~$\mu$Hz to have a relative freedom for this parameter.

For parameter $\Delta\Pi_1$, we chose a normal-uniform prior centered on a value $\Delta\Pi_{1,0}$, with a spread of 14~s, setting $\sigma_\mathrm{max}=$ 2~s and $\sigma_\mathrm{min} = 10$~s, thus allowing for lower and higher values of $\Delta\Pi_1$. The value of $\Delta\Pi_{1,0}$ was obtained from the degeneracy sequence, using the knowledge of $\Delta\nu$. In the case where the electron gas in the core of the star is degenerate, stars of mass $M$ coalesce on a sequence in the ($\Delta\nu$, $\Delta\Pi_1$) plane, referred to as the degeneracy sequence \citep{deheuvels_Seismic_2022}. In the case where the core is not fully degenerate, stars are above the degeneracy sequence ($\Delta\Pi_1>\Delta\Pi_{1,0}$). There also exist cases where $\Delta\Pi_1<\Delta\Pi_{1,0}$, which are attributed to mass transfer \citep{deheuvels_Seismic_2022}. These cases justify the asymmetric spread of the prior of $\Delta\Pi_1$.

For the rotation, we chose priors in agreement with the measurements for core and envelope rotation in \cite{li_asteroseismic_2024}. For the magnetic shift $\nu_\mathrm{mag}$, we chose as an uninformative prior a modified Jeffrey prior with bounds 0.05~$\mu$Hz and 10~$\mu$Hz, which is scale-invariant. Then, because parameter $a$ only takes values between $-1/2$ and $1$, its uniform prior covers all of possible values. 
Lastly, the model error $\sigma_g$ on $g$ mode frequency has a uniform prior between $0$ and $0.1\,\mu$Hz and the model error on $p$ modes is normal-uniform between $0$ and $0.3\,\mu$Hz. These values were chosen based on model errors computed for red giants (see \citealt{li_asteroseismic_2024}). It is also useful to mention that in the case of a star showing a doublet (i.e., two ridges in the stretched échelle diagram) or a singlet (i.e., one ridge), we cannot measure both $\nu_\mathrm{mag}$ and $a$, so we fit the product $\tilde{\nu}_\mathrm{mag}=f_m(a)\nu_\mathrm{mag}$, as mentioned in Sect.~\ref{Model}. In the case of singlets, rotational parameters are not fitted.

\begin{table}
\caption{Choice of priors. In the last column, ``U'' stands for uniform distribution, ``UP'' for uniform periodic, ``UG'' for normal (or Gaussian) uniform and ``J'' for modified Jeffrey prior.} 
\label{table:prior}      
\centering
\small
\begin{tabular}{c c c c c c}
    \hline\hline
        Parameter & $x_\mathrm{min}$ $x_\mathrm{max}$ & $\sigma_\mathrm{min}$ & $\sigma_\mathrm{max}$ & Law \\
        \hline
         
         $\Delta\nu$ ($\mu$Hz) & $\Delta\nu_0-3$&$\Delta\nu_0+3$ & & & U\\
         $\epsilon_p$ & $0.5$ & $1.5$ & & & UP\\
         $\alpha$ & $-0.05$ & $0.05$ & $0.005$ & $0.005$& UG \\
         $d_{01}$ ($\mu$Hz) & $-0.2$ & $0.2$ & $0.1$ & $0.1$ & UG \\
         
         $\epsilon_g$ & $-0.2$ &$0.8$& & & UP \\
         $\Delta\Pi_1$ (s) & ${\Delta\Pi_1}_0 -8$ &${\Delta\Pi_1}_0+6$&$2.5$&$10$&UG\\
         
         $q$ & $0$ & $0.25$ & & & U \\
         
         $\nu_{R,g}$ ($\mu$Hz) & $-0.05$ & $2.00$ & $0.00$ & $0.10$& UG \\
         $\nu_{R,p}$ ($\mu$Hz)& $-0.20$ & $0.20$ & $0.10$ & $0.10$& UG \\
         
         $\nu_\mathrm{mag}$ ($\mu$Hz)& $5~10^{-2}$ & $10$ &  &  & J \\
         $a$ & $-0.5$ & $1.0$ & & & U\\  
         $\sigma_g$ ($\mu$Hz)& $0$& $0.1$& & & U \\
         $ \sigma_p$ ($\mu$Hz)&$0$&$0.3$&$0$ & $0.1$ & UG\\
         \hline
    \end{tabular}
\end{table}

To sample the posterior distributions, we used the Asteroseismic Bayesian Inference by MCMC (ABIM) code that we have developed. It is a Fortran code parallelised with OpenMP directives. Sampling is performed with a Markov Chain Monte Carlo method implementing the \emph{stretched move} algorithm proposed by \citet{Goodman_Weare_2010} and including parallel tempering \citep[e.g.,][]{benomar_solarlike_2009}, which improved the exploration of parameter spaces containing numerous local maxima. We typically sample the posterior distributions with 20 parallel chains and 300 walkers for each chain. The initial positions of the walkers are randomly drawn from the prior distributions. The walkers are iterated over 13\,000 steps, with the 3\,000 first steps discarded as burn-in to ensure that chains are stationary.

After the sampling process, we took  care in assessing the completeness of the sampling for every parameter. One way to do this is to observe the resulting corner plots of the distributions. We also checked the correlation times, which are defined as the time in steps necessary for the autocorrelation function of the chain to diminish by a factor of $e$. A high correlation time $\tau_c$ means that at least $\tau_c$ consecutive samples are highly correlated, which signals a bad sampling of the parameter space. In our case, a typical value for a thorough sampling is around $\tau_c \approx 30$, which ensures to get about 100\,000 uncorrelated sampling points. 

\begin{figure}
    \centering
    \includegraphics[width=\hsize]{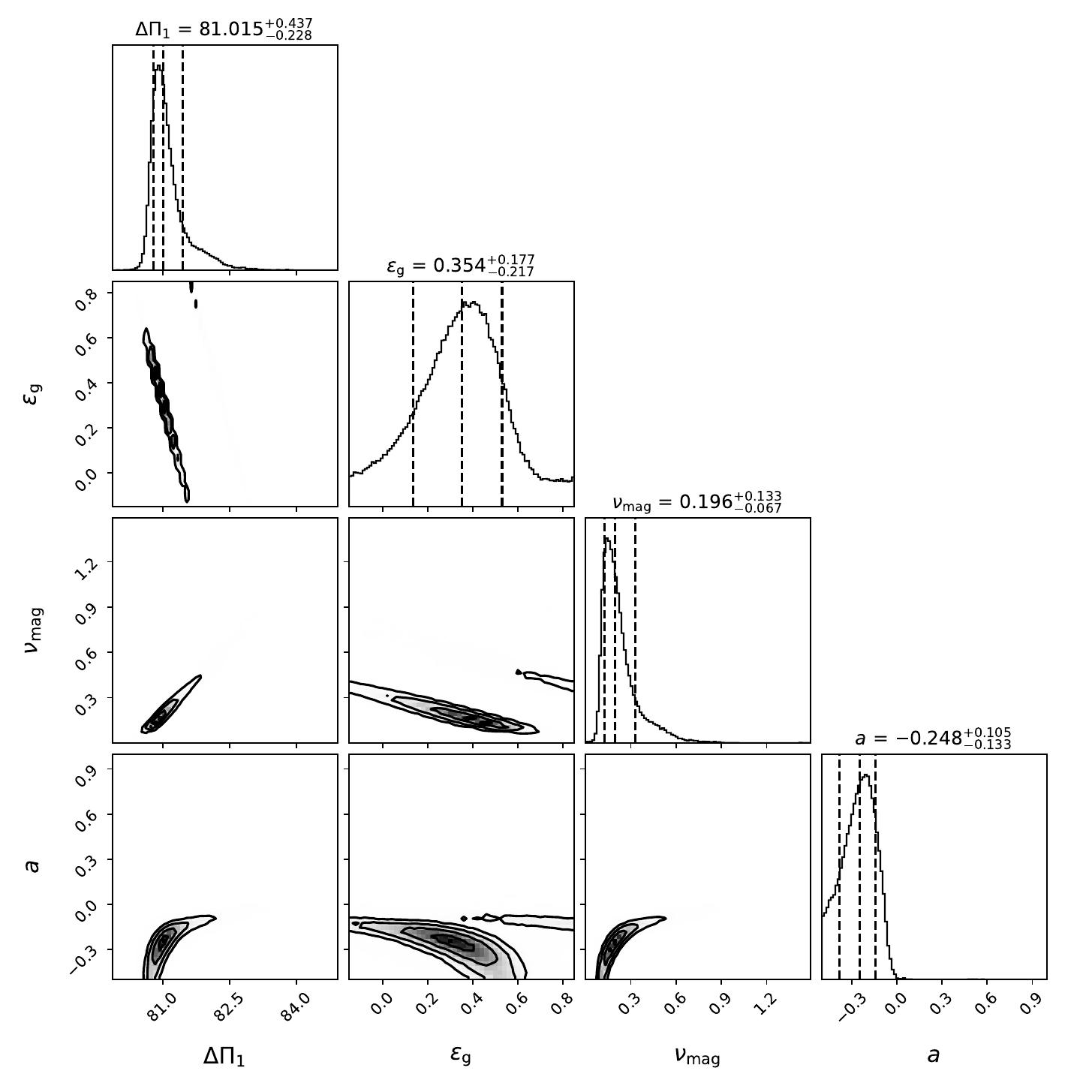}
    \caption{Corner plot of KIC4660930, reduced to only show $g$-mode and magnetic parameters}
    \label{fig:cor-kic4660930}
\end{figure}

An example for a favourable detection of magnetic field can be seen in Fig. \ref{fig:cor-kic4660930},  showing corner plot reduced to magnetic and $g$-mode parameters.
\begin{figure*}
    \resizebox{\hsize}{!}{\includegraphics{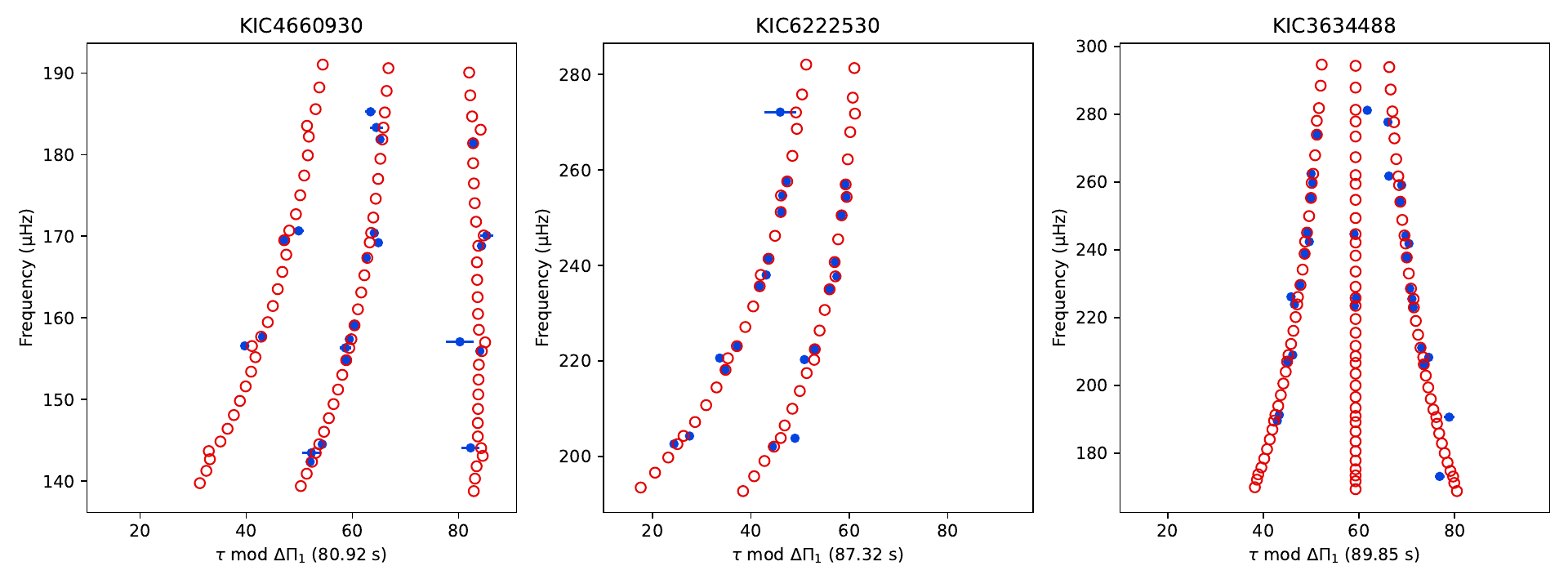}}
    \caption{Stretched \'echelle diagrams of KIC~4660930, KIC~6222530 and KIC~3634488. Blue dots represent detected frequency and red circles represent the maximum a posterior (MAP) model of the determination.}
    \label{fig:echelle3}
\end{figure*}
We can clearly see that the posterior distribution of $\nu_\mathrm{mag}$ drops to zero for low values of $\nu_\mathrm{mag}$, which are strongly excluded because of the negative asymmetry detected in this star. We can also notice that the distribution of $\epsilon_g$ spans across the interval of its prior and has large error-bars. This is most likely due to the small number of detected $g$ modes which are directly used in the measure of $\epsilon_g$.
A representation of the stretched échelle diagram (thereafter s.e.d) of KIC~4660930 can be seen in Fig.~\ref{fig:echelle3} along with the two following examples. There, we can observe the clear asymmetry of the component of the triplet, as well as a significant shift to the left of all three mode components at lower frequency due to the magnetic field. Another example, this time of a doublet, is shown in Fig.~\ref{fig:cor-kic6222503}.
\begin{figure}
    \centering
    \includegraphics[width=\hsize]{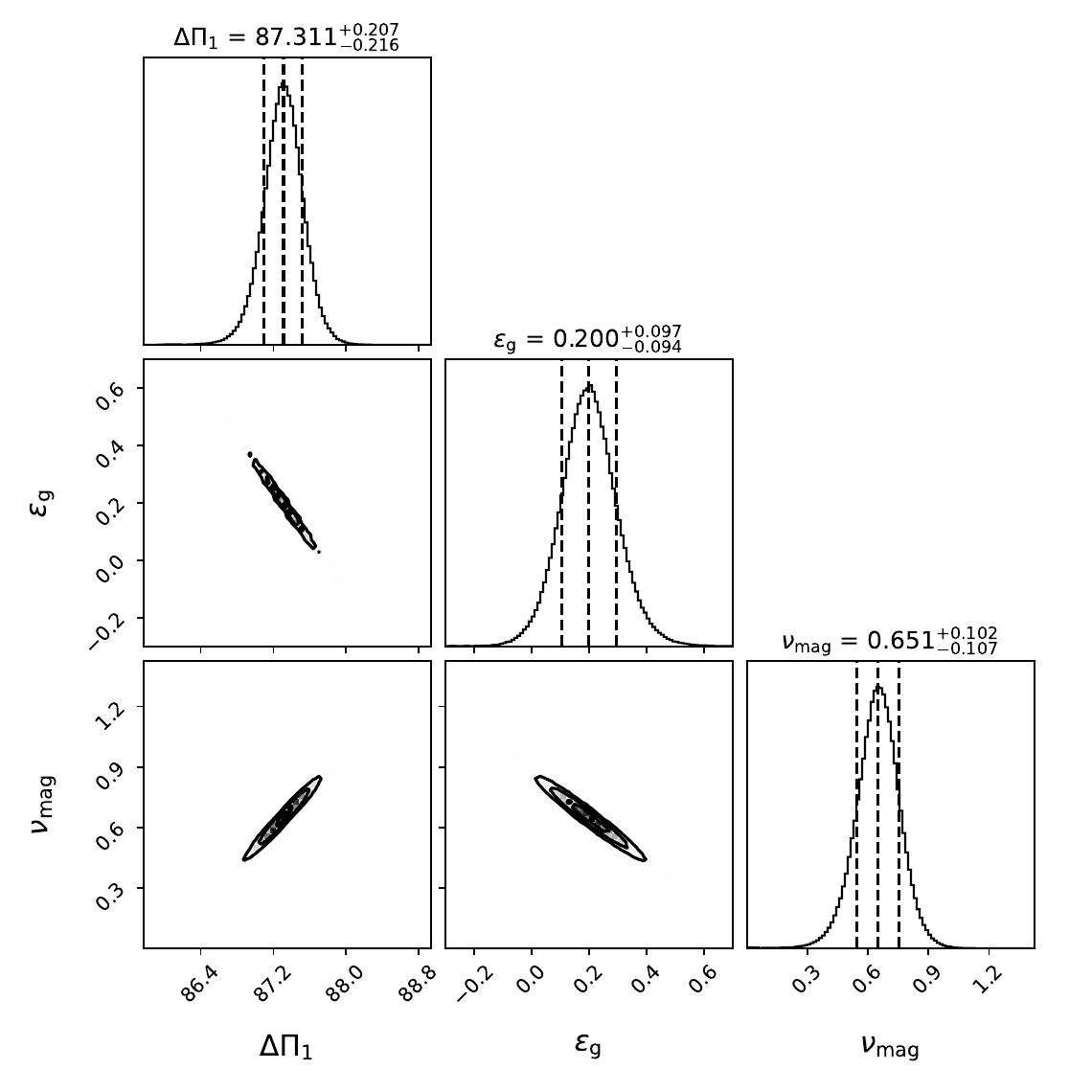}
    \caption{Corner plot of KIC6222530, reduced to show only $g$ mode parameters and the magnetic shift $\nu_\mathrm{mag}$}
    \label{fig:cor-kic6222503}
\end{figure}
Here, the asymmetry parameter cannot be measured because of the lack of the $m=0$ mode component. Nonetheless, we can clearly observe the distribution of $\nu_\mathrm{mag}$ to have a non zero value because of the strong curvature in the s.e.d. A last example is shown in Fig.~\ref{fig:cor-kic3634488}, where the detection is not favourable.
\begin{figure}
    \centering
    \includegraphics[width=\hsize]{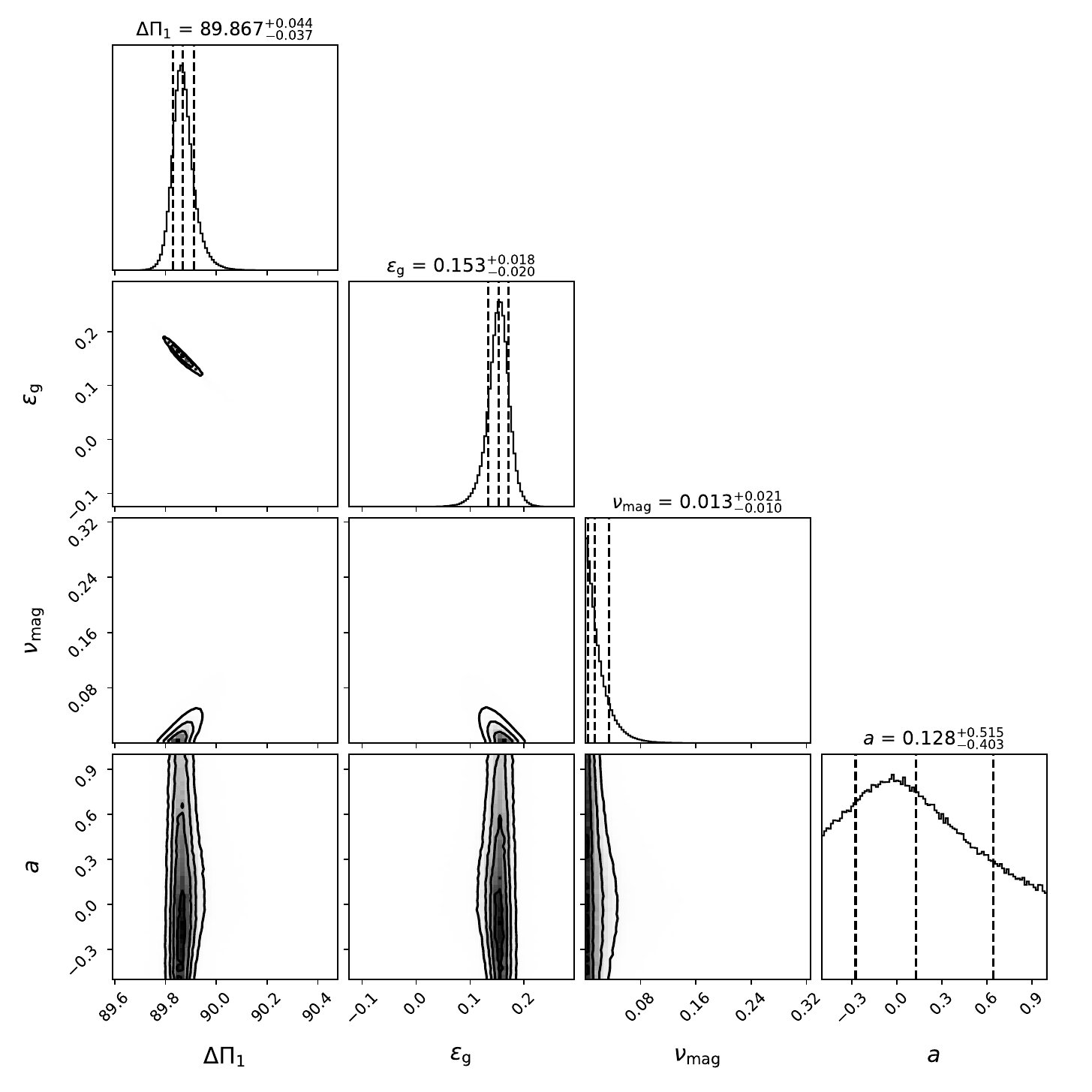}
    \caption{Corner plot of KIC3634488, reduced to show only $g$ mode and magnetic parameters}
    \label{fig:cor-kic3634488}
\end{figure}
Here we see a sharply decreasing distribution for $\nu_\mathrm{mag}$, with its maximum value at $\nu_\mathrm{mag}=0$. We also notice a very large distribution for the asymmetry, almost reproducing the prior but favouring $a=0$. The distribution of $\nu_\mathrm{mag}$ is a strong indicator that this observation is highly compatible with a null magnetic field. This can also be seen in the s.e.d. where the detected frequencies are well explained without magnetic field: the triplet is symmetrical and no discernible shift is observable on the diagram. Among the 218 stars analysed, 13 were dismissed because of a low signal-to-noise ratio and one was removed because of its suppressed dipole modes that could not be analysed. One other star had to be removed from the sample because its $\nu_\mathrm{max}$ was too close to the Nyquist frequency, causing significant aliasing at high frequency. 

\subsection{Probability of magnetic field detection}

In this section we explain the different steps leading to the measurement of core magnetic fields for the cases where a field was significantly detected.

    \label{modelc}

    Assessing the probability of detection of a magnetic field using the posterior distribution can be ambiguous.
    Stars for which non-magnetic models (i.e., with $\nu_\mathrm{mag} \sim 0 \,\mathrm{\mu Hz}$)
    have been seldom visited during the MCMC sampling are good candidates for magnetic field detection, but we need to quantify this. For this, we compare a non-magnetic model $M_0$, where $\nu_\mathrm{mag}$ is fixed to 0, and a magnetic model $M_1$ by computing the odd ratio of $M_0$ over $M_1$ expressed as
    \begin{equation}
        O_{0,1}\; = \;\frac{p(M_0|\Vec{y})}{p(M_1|\Vec{y})}\;=\;\frac{p_I(M_0)}{p_I(M_1)}\frac{p(\Vec{y}|M_0)}{p(\Vec{y}|M_1)}\;, \label{odds}
    \end{equation}
    where $\Vec{y}$ represents the data set, $p(M|\Vec{y})$ represents the probability of the model given the data, $p_I(M)$ is the model prior and $p(\Vec{y}|M)$ represents the global likelihood, or evidence. Here we consider that both models are equiprobable. This implies that the first factor of Eq.~\ref{odds} cancels, leaving what is known as the Bayes factor $B_{0,1}$. To estimate $B_{0,1}$, we computed the global likelihood for models $M_1$ and $M_0$ by taking advantage of properties on the parallel tempering scheme we used \citep[as done, e.g., by][]{benomar_solarlike_2009}. However, in this particular case, we can simplify the computation by taking advantage of  $M_0$ being included in $M_1$ ($M_0$ is indeed a special case of $M_1$ with $\nu_\mathrm{mag}=0$).
    
    We consider that in $M_0$ and $M_1$, all parameters except $\nu_\mathrm{mag}$ constitute nuisance parameters and are represented by a vector $\mathbf{a}$. These parameters share the same support $\mathcal{A}$ and the same priors $p_{I,a}$ for models $M_0$ and $M_1$. In addition, the support of $\nu_\mathrm{mag}$ for model $M_0$ (which corresponds to $\{0\}$) is included in the support of model $M_1$. In this case, the expression of $B_{0,1}$ can be simplified: using Bayes law and strategically marginalising the posterior distribution over vector $\mathbf{a}$, accounting for the independence of the priors on $\mathbf{a}$ and $\nu_\mathrm{mag}$, $B_{0,1}$ can be expressed as:
    \begin{equation}
        B_{0,1} \;=\; \int_\mathcal{B}\frac{p_{I,\nu_\mathrm{mag},0}({\nu_\mathrm{mag}})}{p_{I,\nu_\mathrm{mag},1}(\nu_\mathrm{mag})}p_{\nu_\mathrm{mag}}(\nu_\mathrm{mag})\hbox{d}\nu_\mathrm{mag}\;,
        \label{bayes1}
    \end{equation}
    where $\mathcal{B}$ is the support of $\nu_\mathrm{mag}$ and $p_{\nu_\mathrm{mag}}(\nu_\mathrm{mag})$ is the marginalised posterior distribution. In our case, $\nu_\mathrm{mag}$ is fixed to 0 in model $M_0$. This means that $p_{I,\nu_\mathrm{mag},0}(\nu_\mathrm{mag}) $ is a Dirac distribution at $\nu_\mathrm{mag} =0$, reducing (\ref{bayes1}) to:
    \begin{equation}
        B_{0,1} \;=\; \frac{p_{\nu_\mathrm{mag}}(0)}{p_{I,\nu_\mathrm{mag},1}(0)}\;,
    \end{equation}
    because $p_{I,\nu_\mathrm{mag},1}(0)$ is non-zero.
    Using the first equality in Eq.~\ref{odds}, we can express the probability of a favourable detection as:
    \begin{equation}
        P_1 = \frac{1}{1+ B_{0,1}}\;.
    \end{equation}
    
    In practice, determining $p_{\nu_\mathrm{mag}}(0)$ can be problematic. Ideally, it could be best determined for a continuous distribution, where $p_{\nu_\mathrm{mag}}$ would be perfectly described by a function $p(\nu_\mathrm{mag})$. However, because of the nature of our sampling, $p_{\nu_\mathrm{mag}}$ is a discrete distribution. In order to determine $p_{\nu_\mathrm{mag}}(0)$, we can simply extrapolate the function $p(\nu_\mathrm{mag})$ using only the first bin of $p_{\nu_\mathrm{mag}}$. By varying the bin width $h$, we can use a Taylor series expansion to construct a system of equations to solve for $p(0)$, which is effectively $p_{\nu_\mathrm{mag}}(0)$. This method is further detailed in Appendix~\ref{matrix}. While this method is quite reliable when $p_{\nu_\mathrm{mag}}$ has a non-zero value for $\nu_\mathrm{mag}=0$, it is not the case when very few models were explored around zero (in the case of a very clear detection of a magnetic field). In that case, the values can at best be 0 or very low, as expected, and at worst can yield negative values. In the latter case, $p_{\nu_\mathrm{mag}}(0)$ is simply set to zero.

    We chose to accept a model as favourable if $P_0\,>\, 85\%$, with $85\%$ being an arbitrary percentage. Among the 207 red giants that were investigated, we find \Nstar{} stars with a favourable probability of possessing a magnetic core, the list of which is given in Table \ref{magtab}. This number drops to 18 for a 90\% probability and to 15 for a 98\% probability.
    
\subsection{Magnetic field intensity}
\defcitealias{Paxton2013}{2013}
\defcitealias{Paxton2015}{2015}

\label{field intensity}
The measurement of the field intensity requires the knowledge of the core structure, according to Eq.~\ref{mag1}. We can obtain the value of integral $\mathcal{I}$ by computing an evolution model that matches the seismic parameters that we measured using mixed modes. We computed a grid of models using the MESA 10108 release (\citealt{paxton_modules_2011}, \citetalias{Paxton2013}, \citetalias{Paxton2015}) in which we searched for the best fitting model.
The EOS used is the OPAL \citep{Rogers2002} EOS. Radiative opacities are primarily from OPAL \citep{Iglesias1993,Iglesias1996}, with low-temperature data from \citet{Ferguson2005}. The used nuclear reaction rates are from NACRE \citep{Angulo1999}.
This grid contains $324$ models, varying in mass ($0.9$ to $2 \,M_\mathSun$ with steps of $0.1\, M_\mathSun$), metallicity ($-0.4$ to $0.4$~dex with steps of 0.1~dex) and overshoot parameter ($0$, $0.1$ and $0.2$). The best fitting model is determined by calculating $\chi^2$, taking into account $\nu_\mathrm{max}$, $\Delta\nu$ and $\Delta\Pi_1$ such that :

\begin{equation}
    \chi^2 = \frac{(\Delta\nu^\mathrm{mod}-\Delta\nu^\mathrm{obs})^2}{\sigma_{\Delta\nu}^2} + \frac{(\Delta\Pi^\mathrm{mod}_1-\Delta\Pi_1^\mathrm{obs})^2}{\sigma_{\Delta\Pi_1}^2}+\frac{(\nu_\mathrm{max}^\mathrm{mod}-\nu_\mathrm{max}^\mathrm{obs})^2}{\sigma_{\nu_\mathrm{max}}^2}\;,
    \label{chi2}
\end{equation}
where ``mod'' refers to the value in the model grid, and ``obs'' refers to the median and standard deviation of the posterior distribution for the corresponding parameter. Fitting for $\Delta\Pi_1$ ensures that the core structure is well reproduced and adding $\Delta\nu$ and $\nu_\mathrm{max}$ as constraints ensures that the mass of the model corresponds to the one computed using asteroseismic scaling relations. We managed to find good statistical fits to all our magnetic giants, except for one star, KIC~4350501. This star lies below the degeneracy sequence in the ($\Delta\nu,\Delta\Pi_1$) plane, which is impossible for single-star evolution. This star was already identified by  \cite{deheuvels_Seismic_2022} among stars that are below the degeneracy sequence, and the authors attributed this phenomenon to mass transfer occurring during the post-main-sequence. This explains why no model of our grid can satisfactorily fit KIC~4350501. Mass transfer modified the amount of mass stored in the envelope, but the structure of the core is unchanged. For this star, we thus computed a $\chi^2$ using only the measured $\Delta\Pi_1$ to reproduce the core structure. We recall that when only $m=0$ or $m=\pm1$ are visible, the asymmetry parameter $a$ is not measurable, so that we can only determine intervals for the value of $\langle B_r^2\rangle$. In the case of a singlet, the lower bound of the field intensity is $\langle B_r^2\rangle_\mathrm{min} = \frac{2}{3}\tilde{B_r^2}$, where

\begin{equation}
    \tilde{B_r^2} = \frac{\tilde{\nu}_\mathrm{mag}2\pi\mu_0\omega_\mathrm{max}^3}{\mathcal{I}}\,,
\end{equation}
whereas for doublets, the field strength ranges from $\langle B_r^2\rangle_\mathrm{min} = \frac{2}{3}\tilde{B_r^2}$ to $\langle B_r^2\rangle_\mathrm{max} = \frac{4}{3}\tilde{B_r^2}$.
Using Eq.~\ref{mag1}, we could measure the field intensity for the favourable detections. These values, given in Table \ref{magtab}, range from 34 kG to 227 kG, which is compatible with the values found in \citet{li_magnetic_2022,li_internal_2023,hatt_asteroseismic_2024}. The error on the field intensity was estimated by scaling the distribution of $\nu_\mathrm{mag}$ according to Eq. \ref{mag1}, and measuring the standard deviation.  We can notice that the majority of the detections are below the critical field $B_c$, above which magneto-gravity waves no longer propagate in the core \citep{fuller_asteroseismology_2015,stello_prevalence_2016,rui_gravity_2023}. The ratio between field intensity and critical field intensity $B/B_c$ varies between $0.18 $ and $ 0.68$, with an exception for KIC12168062 where $B/B_c = 1.7$. On that star, one can observe that low-frequency modes are suppressed, while higher-frequency modes retain a mixed mode behaviour. This star is further discussed in Sect. \ref{sect_suppression}.

\begin{table*}[ht!]
\caption{Mass, large separations, core rotation, magnetic shifts, asymmetries and field strength for every favourable detection.} 
\label{magtab} 
\centering 
\begin{tabular}{ c c c c c c c c c } 
\hline \hline
  KIC & $ M $($M_{\mathSun}$) & $\Delta \Pi_1(s)$ & $\Delta \nu $($\mu$Hz)  & $\nu_{R,g} $($\mu$Hz) & $\tilde{\nu}_{mag} $($\mu$Hz) & $a$ & $\frac{\langle{B_r}^2 \rangle^{1/2}}{B_{c,min}}$ & $\langle{B_r}^2 \rangle^{1/2}$(kG) \\ [0.5ex] 
\hline 
2584478 & $ 1.32 \pm 0.06 $ & $ 81.894^{+0.171}_{-0.153}$ & $ 13.947^{+0.013}_{-0.013} $ & $0.538^{+0.010}_{-0.011} $ & $ 0.186^{+0.062}_{-0.053}$ & $0.425^{+0.221}_{-0.130}$ & 0.26 &  $ 106 \pm 17 $ \\ [0.5ex] 
 4066278 & $ 1.44 \pm 0.07 $ & $ 75.758^{+0.140}_{-0.807} $ & $ 10.159^{+0.016}_{-0.016} $ &  & $ 0.732^{+0.029}_{-0.341}$ & S & 0.63 & $ > 73 \pm 9 $\\ [0.5ex] 
 4077249 & $ 1.32 \pm 0.06 $ & $ 79.610^{+0.068}_{-0.046}$ & $ 12.484^{+0.033}_{-0.033} $ & $0.829^{+0.014}_{-0.013} $ & $ 0.077^{+0.023}_{-0.014}$ & $0.821^{+0.128}_{-0.198}$ & 0.18 &  $ 52 \pm 7 $ \\ [0.5ex] 
 4350501 & $ 1.55 \pm 0.08 $ & $ 70.041^{+0.198}_{-0.081}$ & $ 11.137^{+0.014}_{-0.014} $ & $0.142^{+0.010}_{-0.005} $ & $ 0.235^{+0.077}_{-0.032}$ & D & 0.33 & $ [39;55] \pm 23 $\\ [0.5ex] 
 4660930 & $ 1.60 \pm 0.10 $ & $ 80.923^{+0.476}_{-0.232}$ & $ 13.215^{+0.033}_{-0.042} $ & $0.999^{+0.059}_{-0.065} $ & $ 0.175^{+0.144}_{-0.069}$ & $-0.225^{+0.092}_{-0.126}$ & 0.26 &  $ 90 \pm 34 $ \\ [0.5ex] 
 4771876 & $ 1.10 \pm 0.06 $ & $ 86.711^{+0.147}_{-0.143} $ & $ 15.225^{+0.023}_{-0.023} $ &  & $ 0.237^{+0.050}_{-0.049}$ & S & 0.29 & $ > 97 \pm 11 $\\ [0.5ex] 
 4995788 & $ 1.19 \pm 0.06 $ & $ 84.658^{+0.094}_{-0.098} $ & $ 13.524^{+0.014}_{-0.014} $ &  & $ 0.145^{+0.037}_{-0.039}$ & S & 0.24 & $ > 69 \pm 10 $\\ [0.5ex] 
 5287034 & $ 1.25 \pm 0.06 $ & $ 79.090^{+0.101}_{-0.090}$ & $ 12.219^{+0.012}_{-0.012} $ & $0.842^{+0.009}_{-0.008} $ & $ 0.106^{+0.034}_{-0.029}$ & $0.460^{+0.224}_{-0.146}$ & 0.21 &  $ 58 \pm 9 $ \\ [0.5ex] 
 6103082 & $ 1.31 \pm 0.07 $ & $ 85.966^{+0.277}_{-0.370}$ & $ 16.368^{+0.025}_{-0.024} $ & $0.616^{+0.049}_{-0.043} $ & $ 0.970^{+0.135}_{-0.175}$ & D & 0.54 & $ [190;269] \pm 39 $\\ [0.5ex] 
 6222530 & $ 1.37 \pm 0.06 $ & $ 87.316^{+0.207}_{-0.216}$ & $ 17.129^{+0.016}_{-0.015} $ & $0.773^{+0.012}_{-0.012} $ & $ 0.656^{+0.102}_{-0.107}$ & D & 0.43 & $ [196;277] \pm 22 $\\ [0.5ex] 
 6389421 & $ 1.13 \pm 0.07 $ & $ 87.254^{+0.157}_{-0.161} $ & $ 15.654^{+0.014}_{-0.014} $ &  & $ 0.421^{+0.069}_{-0.069}$ & S & 0.38 & $ > 139 \pm 13 $\\ [0.5ex] 
 6758291 & $ 1.15 \pm 0.07 $ & $ 89.312^{+0.384}_{-0.355}$ & $ 16.868^{+0.024}_{-0.022} $ & $1.538^{+0.021}_{-0.021} $ & $ 0.942^{+0.158}_{-0.148}$ & D & 0.53 & $ [236;334] \pm 28 $\\ [0.5ex] 
 8350839 & $ 1.26 \pm 0.08 $ & $ 82.307^{+0.266}_{-0.294} $ & $ 12.699^{+0.027}_{-0.028} $ &  & $ 0.205^{+0.079}_{-0.090}$ & S & 0.30 & $ > 56 \pm 13 $\\ [0.5ex] 
 8574234 & $ 1.06 \pm 0.05 $ & $ 78.572^{+0.129}_{-0.139} $ & $ 10.551^{+0.012}_{-0.012} $ &  & $ 0.234^{+0.038}_{-0.040}$ & S & 0.36 & $ > 39 \pm 4 $\\ [0.5ex] 
 8687248 & $ 1.35 \pm 0.07 $ & $ 79.273^{+1.697}_{-1.873} $ & $ 13.156^{+0.088}_{-0.112} $ &  & $ 1.158^{+0.519}_{-0.558}$ & S & 0.68 & $ > 155 \pm 50 $\\ [0.5ex] 
 8868464 & $ 1.43 \pm 0.09 $ & $ 85.365^{+0.208}_{-0.214}$ & $ 15.441^{+0.018}_{-0.018} $ & $0.407^{+0.011}_{-0.013} $ & $ 0.209^{+0.075}_{-0.080}$ & D & 0.26 & $ [80;114] \pm 21 $\\ [0.5ex] 
 9649838 & $ 1.69 \pm 0.11 $ & $ 74.277^{+0.180}_{-0.127}$ & $ 10.461^{+0.016}_{-0.017} $ & $0.836^{+0.018}_{-0.018} $ & $ 0.149^{+0.060}_{-0.048}$ & $0.464^{+0.238}_{-0.129}$ & 0.28 &  $ 47 \pm 17 $ \\ [0.5ex] 
 10593078 & $ 1.52 \pm 0.08 $ & $ 83.202^{+0.067}_{-0.067} $ & $ 15.307^{+0.015}_{-0.015} $ &  & $ 0.320^{+0.034}_{-0.034}$ & S & 0.32 & $ > 114 \pm 6 $\\ [0.5ex] 
 10784759 & $ 1.64 \pm 0.10 $ & $ 87.726^{+0.198}_{-0.106}$ & $ 15.711^{+0.011}_{-0.011} $ & $1.131^{+0.012}_{-0.011} $ & $ 0.137^{+0.086}_{-0.042}$ & $0.233^{+0.332}_{-0.129}$ & 0.20 &  $ 151 \pm 42 $ \\ [0.5ex] 
 11087371 & $ 1.41 \pm 0.08 $ & $ 78.097^{+0.030}_{-0.027} $ & $ 11.546^{+0.010}_{-0.010} $ &  & $ 0.231^{+0.011}_{-0.010}$ & S & 0.32 & $ > 52 \pm 1 $\\ [0.5ex] 
 11602793 & $ 1.38 \pm 0.07 $ & $ 76.482^{+0.080}_{-0.080}$ & $ 10.453^{+0.009}_{-0.009} $ & $1.138^{+0.004}_{-0.004} $ & $ 0.068^{+0.024}_{-0.023}$ & D & 0.19 & $ [18;25] \pm 4 $\\ [0.5ex] 
 12008916 & $ 1.47 \pm 0.09 $ & $ 80.162^{+0.060}_{-0.059}$ & $ 12.880^{+0.010}_{-0.010} $ & $0.856^{+0.005}_{-0.005} $ & $ 0.107^{+0.020}_{-0.020}$ & D & 0.21 & $ [35;50] \pm 4 $\\ [0.5ex] 
 12168062 & $ 1.49 \pm 0.09 $ & $ 83.201^{+1.689}_{-1.896} $ & $ 12.445^{+0.159}_{-0.152} $ &  & $ 6.828^{+0.255}_{-0.315}$ & S & 1.72 & $ > 327 \pm 52 $\\ [0.5ex] 
 \hline 
\end{tabular}
\tablefoot{ An "S" or "D" in column "$a$" refers to either \textit{singlet} or \textit{doublet} detection. Brackets indicate lower and upper bounds for the corresponding parameter. Masses are taken from \cite{li_asteroseismic_2024}.}
\end{table*}

\section{Comparison with previous studies}
\label{comparison}
\subsection{Gravity offset}
\begin{figure}
    \centering
    \includegraphics[width=\hsize]{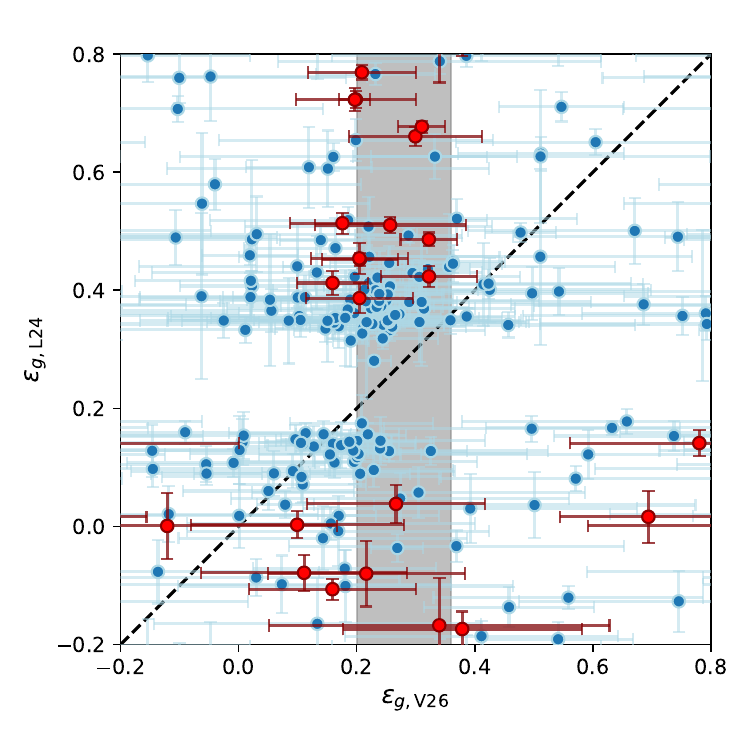}
    \caption{Comparison of $\epsilon_g$ between \cite{li_asteroseismic_2024} and this study. Blue dots represent stars with no detected magnetic field and red dots represent stars with magnetic detections. The grayed out area represents the empirical value of $\epsilon_g$ at $0.28\pm0.08$ \citep{mosser_period_2015}. The black dashed line represents the 1:1 line}
    \label{fig:epsgvsli}
\end{figure}

To build our sample, we chose targets showing abnormal values of $\epsilon_g$, on the ground that the presence of magnetic fields could bias their measurements. To verify this, we plotted the newly measured values of $\epsilon_g$ (taking magnetic fields into account) as a function of the previous measurements (ignoring magnetic fields) in Fig. \ref{fig:epsgvsli}.
We notice that all the magnetic detections in this study have a measured $\epsilon_g$ that is compatible with the value expected for red giants ($0.28 \pm 0.08$). This confirms that for these stars, the abnormal value of $\epsilon_g$ that was obtained in \cite{li_asteroseismic_2024} was indeed caused by the neglect of magnetic fields.
We also found that only 4.8\% of our non-detections are more than $2\sigma$ away from the expected value. This means that almost all non-detections are compatible with the expected value, contrary to what was found in \cite{li_asteroseismic_2024}. This disagreement is caused by the uncertainties in the determination of $\epsilon_g$, which are smaller than the ones we obtained in this study. The difference between the two can be attributed to the way the model error was taken into account. For the stars we studied, we obtained a mean model error of 24~nHz for $g$-modes and 97~nHz for $p$-modes. Comparatively, \cite{li_asteroseismic_2024} used 7 and 14~nHz respectively, which thus seems underestimated.
 
Among the non-detections, we can identify two clumps at $\epsilon_{g,\mathrm{L24}} \approx 0.15$ and $\epsilon_{g,\mathrm{L24}} \approx 0.38$. These stars make up the tails of the distribution of ``normal'' stars in \cite{li_asteroseismic_2024}, which is why we recover values compatible with the expected value in this study. We also see that almost no magnetic detection lies among these clumps, further reinforcing the idea that these stars should be considered as normal. 
The rest of the non-detections are more spread out and have larger error bars. This can be linked to difficulties in fitting the observed frequencies, where we recovered either multiple solutions or flat posteriors, which drives up the width of error bars.

\subsection{Rotation rates}
\label{comprate}
\begin{figure}
    \centering
    \includegraphics[width=\hsize]{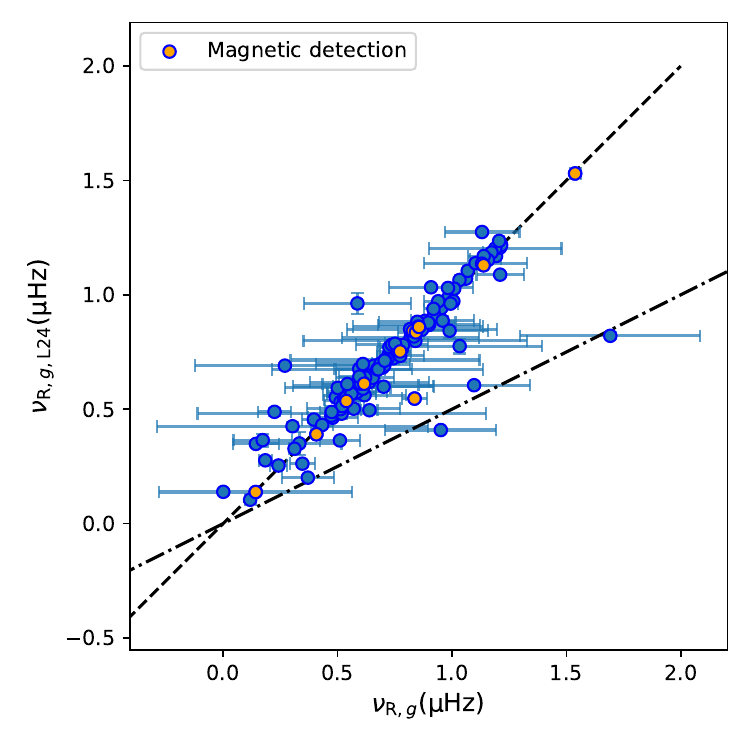}
    \caption{Comparison of core rotation measured in \cite{li_asteroseismic_2024} and in this study. The blue dots represent measurements for non-magnetic stars, and the orange dots represent magnetic stars with $1\sigma$ error bars. The dashed black line represents the 1:1 line and the dashed-dotted line represents the 1:2 line.}
    \label{fig:rotliicig}
\end{figure}

As a result of our fitting procedure, we also obtained estimates of the core and envelope rotation rates for stars with $l=1$ triplets or doublets. To assess the robustness of our method, we show the measured core rotation rates as a function of previous measurements in Fig.~\ref{fig:rotliicig}. 
Most measurements fall on the $1:1$ line, indicating that we generally recover the same rotation rates as in \cite{li_asteroseismic_2024}. We also notice that a small fraction of measurements fall on the $1:2$ line. In these stars, we detected a previously undetected component, effectively doubling the rotation rate. 
Not represented on Fig.~\ref{fig:rotliicig}, are cases where rotation was discovered in this study (\citealt{li_asteroseismic_2024} had detected a singlet, but we detect a doublet or a triplet), concerning 16 stars. We can also see that magnetic stars do not follow a particular tendency on this figure.
We were also able to compare core rotation rates for seven stars in common with \cite{hatt_asteroseismic_2024}. In both studies, no magnetic field was detected in these stars. Our measurements are statistically compatible with those of \cite{hatt_asteroseismic_2024} for these stars.

Concerning envelope rotation, our measurements are compatible with those of \cite{li_asteroseismic_2024}, although our error bars are larger, for reasons explained above.

\section{Discussion}
\label{discussion}

\subsection{Mass distribution of magnetic red giants }

The new detections in this study have an average mass of $1.375 \pm 0.196 M_\mathSun$. With the addition of these  \Nstar{} new stars, the total number of detections is 71 magnetic red giants. This number allows us to begin exploring the distributions of different stellar parameters (e.g., mass, rotation) for these stars and see how they compare to those of red giants in general. The distributions for all well-characterized red giants are taken from the catalogue of \cite{li_asteroseismic_2024}. Since all the studies that searched for magnetic fields in red giants so far have focused on low-luminosity stars, we decided to select only the stars from \cite{li_asteroseismic_2024} for which $\Delta\nu> 10$~$\mu$Hz, for consistency reasons.
 
\begin{figure}
    \centering
    \includegraphics[width=\hsize]{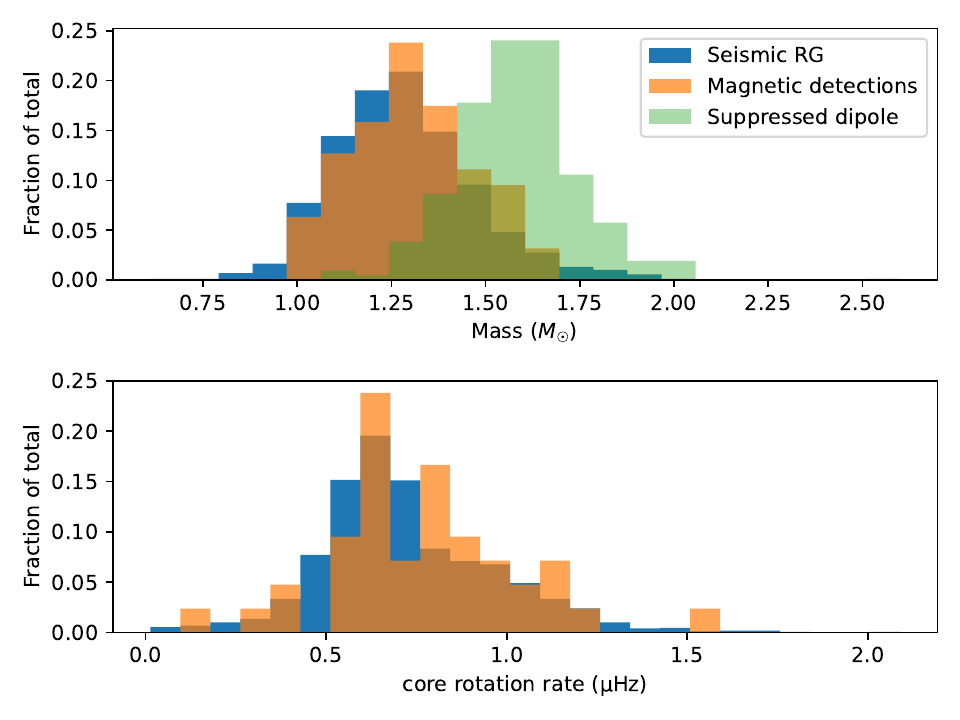}
    \caption{\textbf{Top}: Normalized histogram of stellar mass comprised of data from \cite{hatt_asteroseismic_2024}, \cite{li_asteroseismic_2024}, \cite{deheuvels_strong_2023} and this study. The distribution for magnetic detections is shown in orange and the entire distribution of red giants is represented in blue. The distribution for dipole depressed modes in \cite{stello_prevalence_2016} is shown on green.  
    \textbf{Bottom}: Normalized histograms of the core rotation rate $\nu_\mathrm{R,g}$, comprised of data from the same previous studies. The colours are the same as above. }
    \label{fig:massrothist}
\end{figure}

By plotting the histograms of the mass distribution in Fig. \ref{fig:massrothist}, we find that the two distributions seem quite similar. We can see that the magnetic distribution follows the (much more populous) complete distribution. Computing a Kolmogorov-Smirnov test shows that the two distributions have a 50\% chance of following the same probability law. If this tendency holds up with the growing number of magnetic detections, this would show that magnetic red giants are not peculiar from a stellar mass standpoint. The distribution of magnetic stars and non magnetic stars being similar could also imply that more possible detections still lie in the non detection distribution.

We also represented the distribution for stars showing suppressed dipole modes in \cite{stello_prevalence_2016}, which are suspected of harbouring high-intensity magnetic fields. The stars represented here all have $\Delta\nu>10~\mu$Hz to be coherent with our sample in terms of evolutionary state. We can clearly see the difference between the two distributions, the suppressed-dipole-mode stars having much higher masses. One possible explanation for this discrepancy could be that the two populations are indeed different, and that the fields causing suppressed dipole modes are different from the fields we detect. 
This could also be caused by observational biases. Magnetic fields are directly detectable using seismology if they are strong enough to produce a detectable signature but not stronger than the critical field $B_c$. This gives a range in magnetic field strength for which fields are detectable. If this range becomes narrower with increasing stellar mass, this could explain the relative lack of more massive red giants with detected magnetic fields. This question will be addressed in a later study. The detection threshold decreases as stars ascend the RGB, so the prospects of detecting fields in more massive stars might improve when studying more luminous red giants.

\subsection{Rotation rates}

For the magnetic stars in this study, the core rotation can be measured in cases where either two or three mode components per triplet are visible, as reminded in Sect.~\ref{comprate}. Here, rotation was measured in 12 magnetic stars, which can be added to the ones from \cite{li_magnetic_2022,li_internal_2023} (13 stars) and the ones from \cite{hatt_asteroseismic_2024} (24 stars). Here, we compare the distribution of core rotation rates for magnetic giants with the one made up of the remaining non detections from this study, \cite{li_asteroseismic_2024} and $\cite{hatt_asteroseismic_2024}$. However the size of the first distribution is fairly small, so it should not be over-interpreted.

In the lower panel of Fig.~\ref{fig:massrothist}, we can see that the distribution of core rotation for magnetic stars is very similar to the distribution of core rotation for red giants in general, with a peak around $0.8\,\mu$Hz. This peak contains mostly stars from \citet{hatt_asteroseismic_2024} which can also be seen in the red giant global distribution. A Kolmogorov-Smirnov test using these two distributions indicates that they have high odds of following the same statistical law. It is quite striking that the core rotation rate does not seem to be different for stars that are found to host core magnetic fields. This could of course mean that the detected fields are not responsible for the efficient transport of angular momentum occurring inside red giants. However, this could also mean that magnetic fields are ubiquitous in red giant cores, and that in most cases we have not detected them, either because they produce a seismic signature that is below the detection threshold, or because the detection methods applied so far are ill-suited (for instance for strong non-axisymmetric fields). 

\subsection{Impact of evolution}

\begin{figure}
\centering
    \includegraphics[width=\hsize]{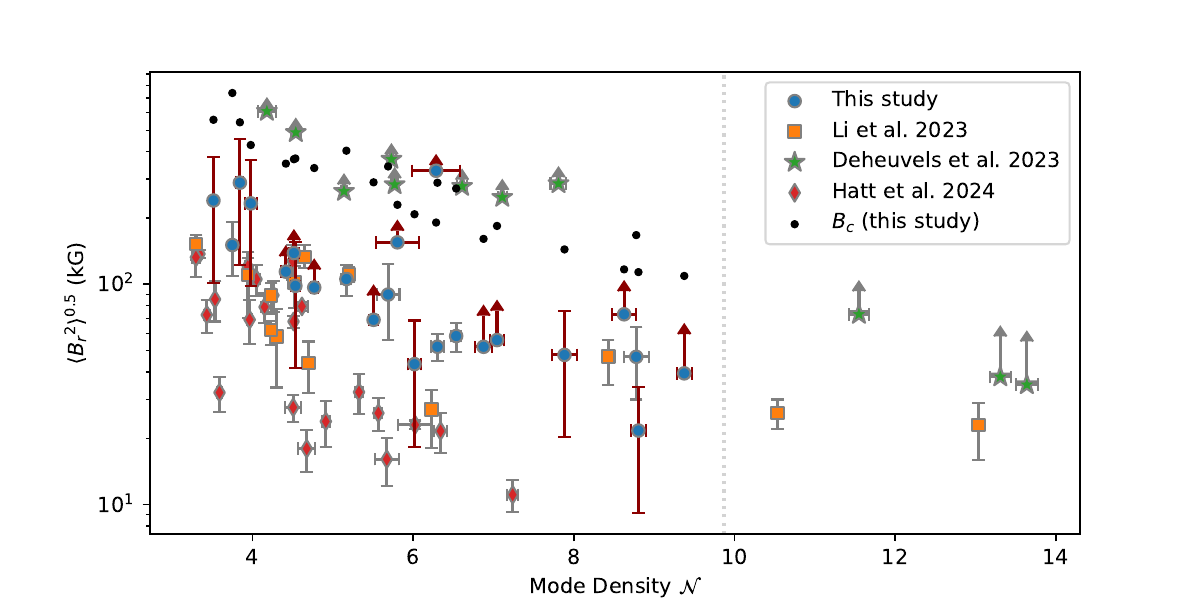}
    \caption{Variation of magnetic field intensity with mode density, used as a proxy for stellar evolution. Four different studies are represented on the plot. The arrows pointing upwards represent lower limits to the field intensity while red, vertical error bars represent the lower and upper limits of the magnetic fields for stars that only show two $l=1$ mode components. The grey dotted line at $\mathcal{N} = 9.87$ represents the upper limit in evolution reachable in this study (calculated  using $\Delta\nu = 10\,\mathrm{\mu Hz}$ and the relevant values for $\Delta\Pi_1$ and $\nu_\mathrm{max}$ obtained from the degeneracy sequence and seismic scaling laws).}
    \label{evol}
\end{figure}

We can also explore the distribution of magnetic stars with respect to evolution. In Fig. \ref{evol}, we plot the magnetic field strength as a function of the mixed mode density $\mathcal{N} = \Delta\nu/(\Delta\Pi_1\,\nu_\mathrm{max}^2)$, which serves as an indicator for evolution \citep[][]{gehan_core_2018}. We confirm previous findings that the intensity of detected fields decreases with evolution (\citealt{deheuvels_strong_2023}, \citealt{li_internal_2023}).

This decreasing trend can seem counter-intuitive. Indeed, if the magnetic flux is conserved, the core contraction as the star ages should cause the field strength to increase. This may be explained through two different phenomena. First, as suggested by \cite{deheuvels_strong_2023}, the field intensity could be increasing with age, but would eventually reach the critical field strength $B_c$ (which decreases with evolution), making the observation of stronger fields impossible. This is supported by the fact that our detections decrease similarly as $B_c$. Another potential explanation can be proposed, assuming that the detected fields are generated in main-sequence convective cores, and survive on the RGB after relaxing into a stable configuration. During this phase, hydrogen is consumed in a shell (the HBS), which extends outwards as the star ages, leaving an inert helium core behind. Eventually, the HBS passes the limit of the main-sequence convective core. Since the seismic signature of magnetic field is mainly sensitive to the HBS, we expect the measured field strength to decrease with evolution when the HBS has left the previously-convective core.

We can also see that the different studies shown cover different parts of the graph, such as \cite{hatt_asteroseismic_2024} covering low-intensity, low-luminosity red giants, or \cite{deheuvels_strong_2023} covering intense fields. This patchwork of detections exhibits the observational biases of these different studies. We can also notice a gap between the near-critical fields and the rest of the measurements. For now, no explanation was given for this gap. This representation also exhibits detection thresholds for field intensity. Indeed, Eq.~\ref{mag1} shows that $\nu_\mathrm{mag}$ varies as ${(\nu_\mathrm{max}})^{-3}$. This means that for low-luminosity red giants with high $\nu_\mathrm{max}$, the magnetic intensity must be very high to observe significant shifts. In a similar way, evolved red giants with low $\nu_\mathrm{max}$ must have field intensities below the critical field to be observable. 

\subsection{Asymmetry parameter $a$}
\begin{figure}
    \centering
    \includegraphics[width=\hsize]{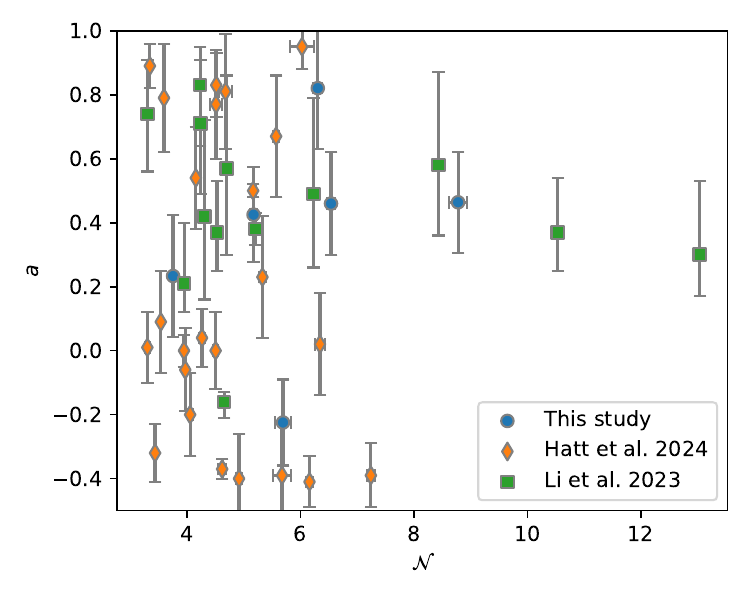}
    \caption{Relation between parameter $a$ and $\mathcal{N}$.}
    \label{fig:asymdpi}
\end{figure}
In this study, we identified six magnetic red giants for which three components per $l=1$ are visible, so that we were able to measure the asymmetry parameter $a$. Adding these measurements to those of previous studies, we now have 44 stars with measured asymmetries \citep[][]{li_internal_2023,hatt_asteroseismic_2024}. Fig.~\ref{fig:asymdpi} shows the relation between the parameter $a$ and the evolution proxy $\mathcal{N}$. We count 7 stars whose asymmetry is compatible with zero, 27 stars with positive asymmetry and 10 with negative asymmetry. The interpretation of tendencies in Fig.
~\ref{fig:asymdpi} is complicated owing to the observational biases of the corresponding studies. \cite{li_internal_2023} specifically searched for non-zero asymmetries. The study of \cite{hatt_asteroseismic_2024} can detect fields with vanishing asymmetry parameter, but they are limited to stars with $\mathcal{N} \lesssim 8$.

We can notice that there is a larger number of stars with positive asymmetry than ones with negative asymmetry. This implies a higher probability for stars to form fields concentrated at the poles rather than ones concentrated at the equators. Also, we clearly see here that a sizeable portion of the detected fields seem incompatible with pure dipolar fields. Indeed, for dipolar fields $a$ ranges from $-0.2$ and 0.4, and we see that a large proportion of stars are still outside these bounds. Thus, considering only dipolar fields as representative configurations for internal magnetic fields is not enough, and different configurations should be considered for accurate representations.

\subsection{Potential non-axisymmetric effects}
\label{nonaxi}
We can assess our assumption on the axisymmetry of the field by computing the parameter $b=2\nu_\mathrm{B}/\nu_\mathrm{R,g}$. \cite{li_magnetic_2022} showed that non-axisymmetric effects can be ignored if $b<1$. We computed the value of $b$ for all our magnetic stars at three different points in the spectrum: for the lowest-frequency detected mode, at $\nu_\mathrm{max}$, and for the highest-frequency detected mode. Since $\nu_B \propto \nu^{-3}$, $b$ also follows this trend. This means that we are more likely to observe non-axisymmetric effects at lower frequencies. We also recall that in cases where only two components are visible, the value of $\nu_{\rm mag}$ is bounded between $(2/3)\tilde{\nu}_\mathrm{mag}$ and $(4/3)\tilde{\nu}_\mathrm{mag}$, hence the same intervals hold for $b$. In Table \ref{btab}, these intervals are represented with brackets, along with the rest of the magnetic detections.

\begin{table}
\caption{Value of parameter $b$ for every favourable detection with more than one $l=1$ mode component}
\label{btab}
\centering 
\begin{tabular}{ c c c c } 
\hline\hline
KIC & $b(\nu_\mathrm{first})$ & $b(\nu_\mathrm{max})$ & $b(\nu_\mathrm{last})$ \\ 
\hline 
2584478 & 1.36 & 0.76 & 0.42 \\ 
4077249 & 1.05 & 0.50 & 0.31 \\ 
4350501 & [3.33;6.65] & [1.93;3.87] & [1.02;2.05] \\ 
4660930 & 0.69 & 0.41 & 0.31 \\ 
5287034 & 1.04 & 0.49 & 0.24 \\ 
6103082 & [0.81;1.61] & [0.44;0.89] & [0.26;0.52] \\ 
6222530 & [0.57;1.13] & [0.35;0.71] & [0.23;0.46] \\ 
6758291 & [0.32;0.64] & [0.18;0.36] & [0.12;0.24] \\ 
8868464 & [1.18;2.36] & [0.67;1.34] & [0.43;0.87] \\ 
9649838 & 1.27 & 0.49 & 0.27 \\ 
10784759 & 0.60 & 0.36 & 0.24 \\ 
11602793 & [0.61;1.21] & [0.24;0.48] & [0.10;0.21] \\ 
12008916 & [0.96;1.92] & [0.32;0.64] & [0.17;0.34] \\ 
\hline 
\end{tabular} 
\end{table} 

We notice that at least eight stars of this sample exhibit conditions in which non-axisymmetric effects could be observed. These effects would manifest by additional components, up to nine in total for dipole modes \citep[see][supplementary information]{loi_Topology_2021,li_magnetic_2022}. In these stars, no additional component appears at low frequency. For now, we thus do not detect signature of non axisymmetric effects for the magnetic field, even though these stars could show some.

\subsection{Mode suppression \label{sect_suppression}}

\begin{figure}
    \centering
    \includegraphics[width=\hsize]{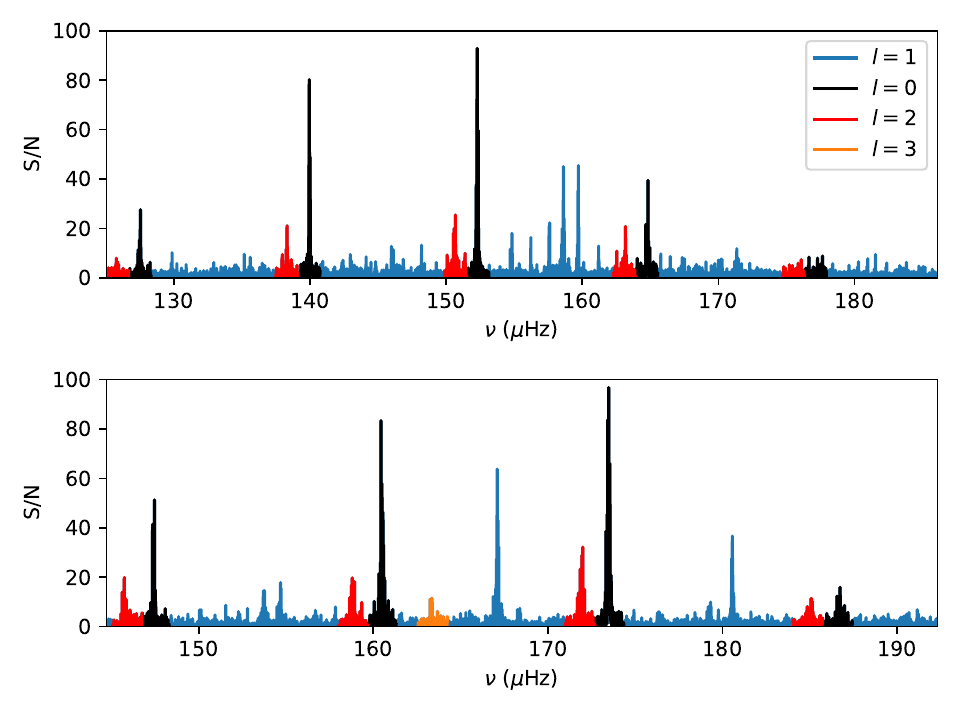}
    \caption{\textbf{Top:} Spectrum of KIC~12168062. \textbf{Bottom:} Spectrum of KIC~8687248. Black areas denote radial modes, red areas denote quadruple modes and orange signals the presence of a visible $l=3$ mode.}
    \label{fig:suppressed1}
\end{figure}

In this study, we found evidence of a magnetic field in two stars showing signs of suppressed dipole modes
The first star, KIC~12168062 (see Sect. \ref{field intensity}), exhibits a strong visibility gradient of dipolar mixed modes. Indeed, the visibility of dipole-modes (measured relatively to the visibility of radial modes) goes from an average of $0.14$ below $\nu_\mathrm{max}$ to $1.49$ above $\nu_\mathrm{max}$. This behaviour is shown in the top panel of Fig.~\ref{fig:suppressed1}, where we can clearly see that dipolar modes are almost absent from the first two large separations, but suddenly appear as they normally would in a star without suppressed dipoles.
The second star, KIC~8687248, was already shown to have suppressed dipole modes in \cite{mosser_dipole_2017}. In the bottom panel of Fig. \ref{fig:suppressed1} we can see how, again, the visibility of dipole modes is lower than that of radial modes, with a mean ratio of the dipolar to radial mode energy of about $0.51$. We can also see that only the most $p$-dominated modes are visible in the spectrum. Despite this fact, we clearly found that these modes follow a mixed mode pattern disturbed by a core magnetic field. 
These two stars also share the two highest $B/B_c$ ratios of our detections, which further links the effect of the critical field to the suppression of modes. This also means that the perturbative approach we used on these two stars likely overestimates the field intensity and a non-perturbative would be required to better treat these stars \citep{rui_asteroseismic_2024}.

\section{Conclusion}

In this study we investigated the use of the gravity offset $\epsilon_g$ as a means to identify magnetic red giants. We worked on 218 stars, considered as abnormal in terms of $\epsilon_g$ in the catalogue of \cite{li_asteroseismic_2024}. We developed a method potentially applicable to all red giants to detect and measure magnetic fields. Since magnetic fields have stronger effects on low-frequency modes, it is crucial to reliably detect these modes. We applied significance tests that we designed in order to optimise the detection probability of the modes, while guaranteeing a low false-alarm probability.
We then adjusted an asymptotic expression of mixed modes to the frequencies of the detected modes, taking magnetic fields and rotation into account. For this purpose, we use a Bayesian inference, which allowed us to explore the posterior distributions for every parameter considered in the model, and to assess the odds of detecting a magnetic field. We then used a grid of stellar model to link the magnetic frequency perturbation to the magnetic field intensity.

Among the 218 stars selected, \Nstar{} exhibit magnetic fields with a high degree of confidence in the detection.  We found intensities for the radial component of the magnetic field ranging from 34~kG to 262~kG, in line with measurements from previous studies. For these stars, we found values of $\epsilon_g$ that are now in agreement with the expected value for red giants, validating the suggestion of \cite{li_magnetic_2022}. For seven stars we placed constraints on the topology of the field, through the asymmetry parameter $a$. We were also able to measure the internal rotation for 12 of those newly detected magnetic giants.

Adding our new measurements to the previous studies measuring core magnetic fields \citep{li_magnetic_2022,li_internal_2023,deheuvels_strong_2023,hatt_asteroseismic_2024}, there are now 71 detections of magnetic red giants, which allowed us to start investigating the distributions of stellar parameters for magnetic giants. We confirmed
the decreasing trend of the radial field component with the evolution of the star, which was already noticed in previous studies. This trend is likely due to observational biases, but could also result from the progression of the hydrogen-burning-shell with evolution, as we proposed in this study. We found that the mass distribution of stars in which we detect magnetic fields is so far indistinguishable from the mass distribution of the complete catalogue of seismic red giants. On the contrary, the masses of red giants with direct field detection are significantly lower than those of stars showing suppressed dipole modes at the same evolutionary stage (\citealt{stello_prevalence_2016}). This discrepancy could mean that the two populations are truly distinct, and that the fields producing mixed-mode suppression differ from the ones we detect directly. It could also be that the range of detectable field strengths (between the detection threshold and the critical field) narrows down as stellar mass increases, leading to less detections for more massive stars. We also found that magnetic red giants have a distribution of core rotation rates that is strikingly similar to that of other red giants. 
This shows that the transport of angular momentum does not act in a different way for stars with detected magnetic fields compared to the rest of red giants. There are two possible interpretations regarding angular momentum redistribution in these stars: either magnetic fields do not have a predominant role in the transport of angular momentum in red giants, or the rest of red giants also host core magnetic fields that have not yet been detected.

This study also brings new insight on field geometry. By studying the distribution of the asymmetry parameter $a$, we showed that positive asymmetries are more frequent than negative ones. Also, many detected fields have measured values of $a$ that are incompatible with a dipolar geometry for $B_r$, so that the assumption of dipolar fields, often made for simplicity, is not relevant. The effects of non-axisymmetry of the fields were also investigated. We showed that at least 8 stars exhibit conditions favourable for the detection of non-axisymmetric effects. However, no such effects were discovered in these stars. 

To conclude, this study represents an effort to bring new detections of core magnetic field in which we have a high degree of confidence. These detections are another step toward building an exhaustive catalogue of magnetic red giants. In this work, we only studied stars with $\Delta\nu<10~\mu$Hz, because more luminous stars have more complex spectra. In a forthcoming study, these stars will be explored using the same method described here. This new method, from the identification of mode frequencies to the measure of magnetic field intensity, is in principle applicable to any oscillation spectrum of red giants, and can already be used on TESS spectra. In the near future, we aim to automate and use this method on the totality of the catalogue of \cite{li_asteroseismic_2024} to look for core magnetic fields. Eventually, this method will be used on PLATO \citep{rauer_plato_2025} spectra also, further expanding the number of detected magnetic red giants.

\begin{acknowledgements}
This work has been supported by CANES, focused on the preparation of the PLATO mission.
This paper includes data collected by the \emph{Kepler} mission and obtained from the MAST data archive at the Space Telescope Science Institute (STScI). Funding
for the \emph{Kepler} mission is provided by the NASA Science Mission Directorate.
STScI is operated by the Association of Universities for Research in Astronomy,
Inc., under NASA contract NAS 5–26555. Software: \texttt{Python} \citep{python}, \texttt{numpy} \citep{oliphant2006guide,harris2020array}, \texttt{matplotlib} \citep{hunter2007matplotlib}, \texttt{scipy} \citep{virtanen2020scipy}, \texttt{corner} \citep{foreman2016corner}, \texttt{astropy} \citep{astropy:2013,astropy:2018,astropy:2022}.
\end{acknowledgements}

\bibliographystyle{aa}
\bibliography{biblio2.bib}

\begin{appendix}
\nolinenumbers
\section{Mode identification method}
\label{extraction}

\begin{figure}
\centering
\includegraphics[width=\hsize]{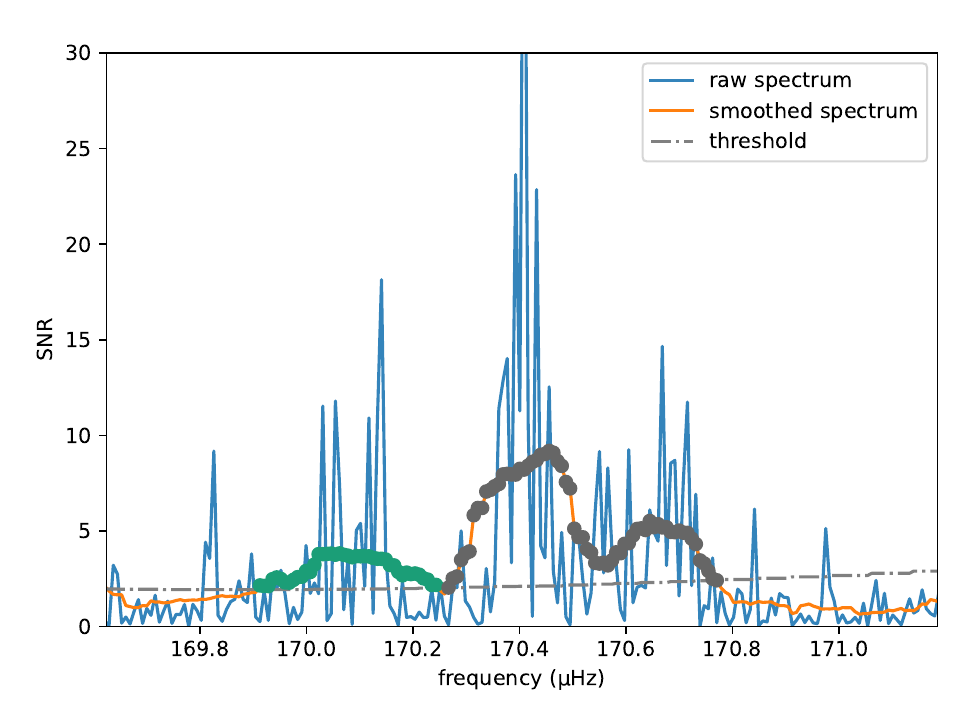}
  \caption{Portion of the PSD of KIC~4660930 showing a mixed $l=1$ mode. The raw spectrum is plotted in blue, the variable smoothed spectrum is plotted in orange, and the dash-dot gray line represents the significance threshold of the smoothed spectrum. The dot markers represent the points of the spectrum that exceed the threshold, and the colours represent the group to which they belong.
          }
     \label{group1}
\end{figure}

In this section, we detail the process used to 
assign a unique frequency and width guess to each significant mode component in the spectrum. As a first step, we apply the variable threshold presented in Sect. \ref{analysis} to the smoothed spectrum. This creates groups of consecutive points containing the peaks of the spectrum that exceed the threshold. Fig.~\ref{group1} shows an example of such groups on a rotational triplet of KIC~4660930. There, we see that the $m=0$ and $m=-1$ components are contained within one single group.

\begin{figure}
\centering
\includegraphics[width=\hsize]{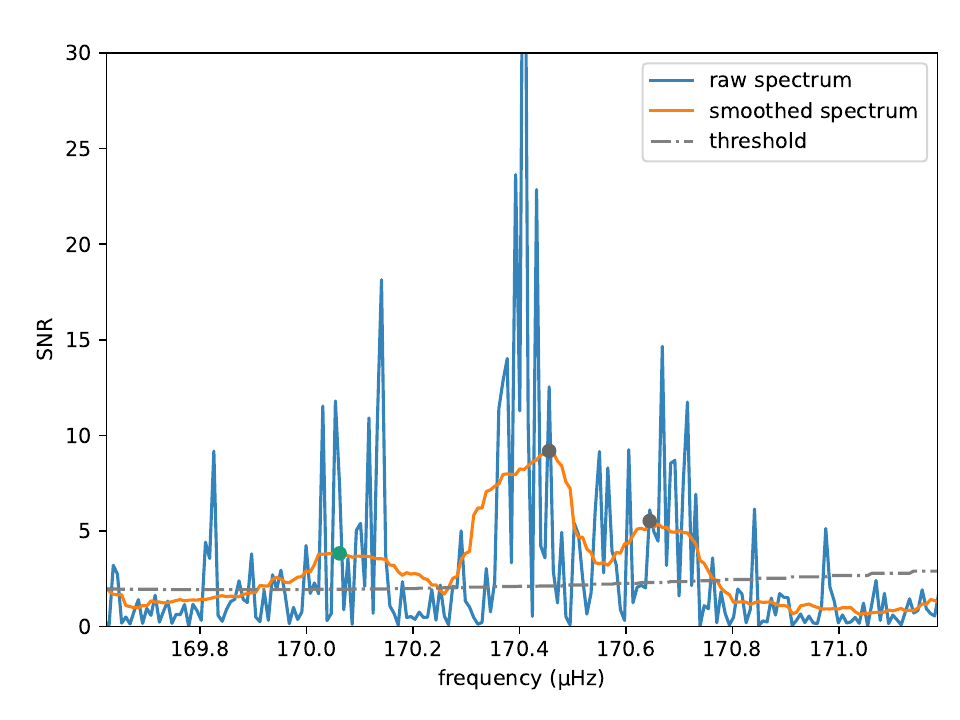}
  \caption{Same as Fig. \ref{group1}, after reducing the number of peaks in each group (see text)}
     \label{group2}
\end{figure}

A second step consists of reducing the number of points in each group. This is done by selecting the local maxima in the group two consecutive times. Local maxima are defined as points where the two neighbouring points are lower than the central one. Selecting these maxima two consecutive times helps reduce the effects of noise in the spectrum, while conserving split modes that are found in one group. The result of this step is shown in Fig. \ref{group2}. We clearly see that this reduces the number of identifications to a manageable number, while preserving both $ m=0$ and $m=1$ components.

\begin{figure}
\centering
\includegraphics[width=\hsize]{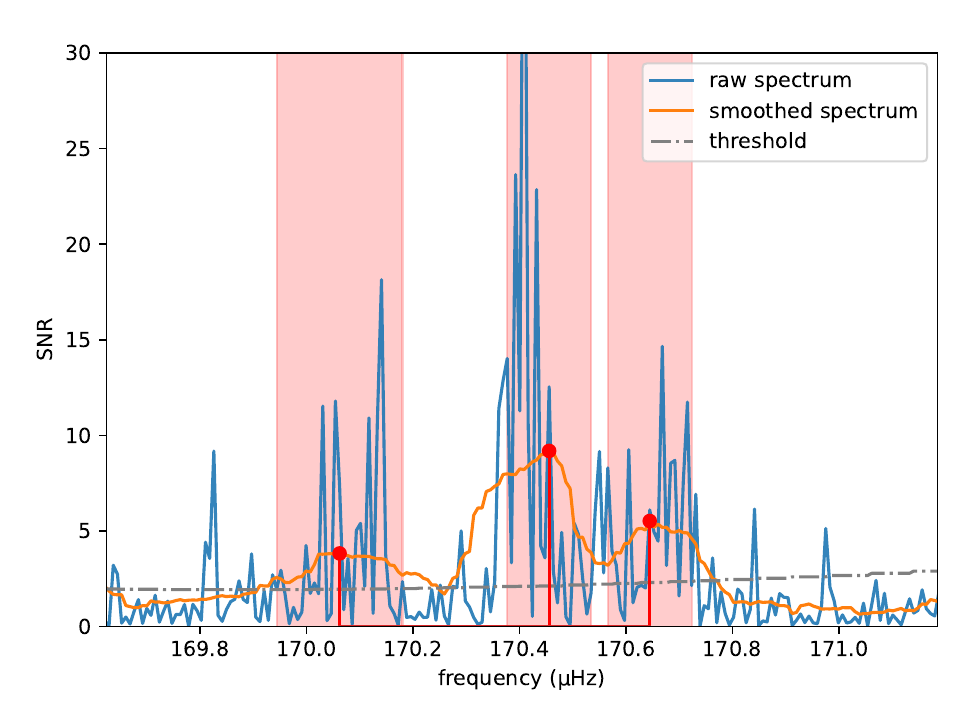}
  \caption{Same as Fig. \ref{group1}. The transparent red zones delimit an area of three times the $l=1$ mode width, estimated using $\zeta$. The red stem plots mark the points that have been selected as guesses for mode frequency.
          }
     \label{group3b}
\end{figure}

\begin{figure}
\centering
\includegraphics[width=\hsize]{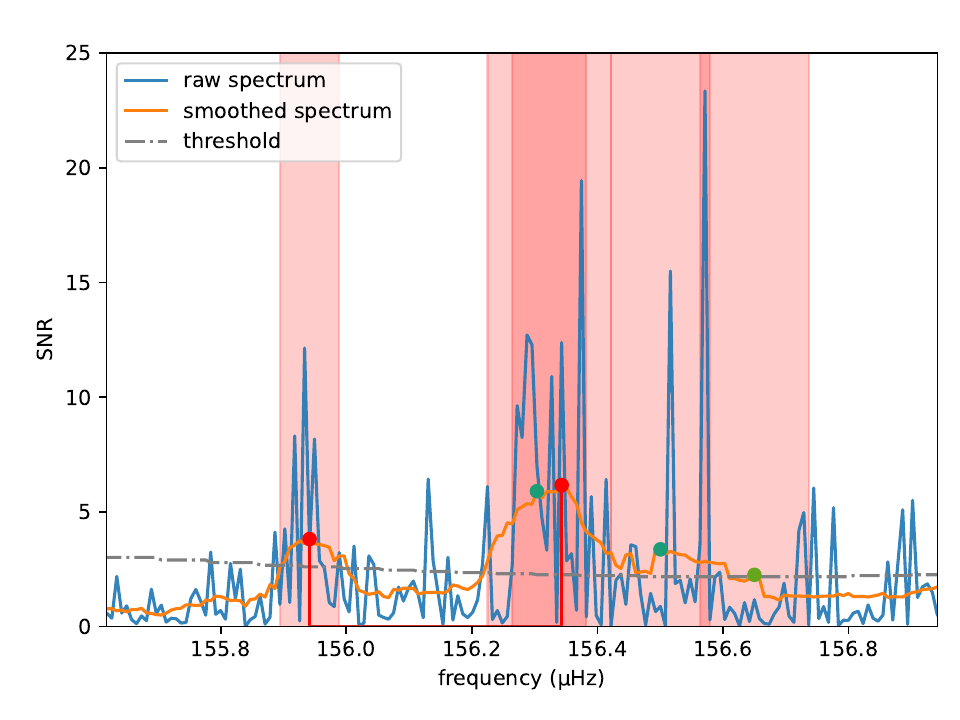}
  \caption{Same as Fig. \ref{group3b} for another p-dominated triplet in the PSD of KIC~4660930. 
          }
     \label{group3}
\end{figure}

\begin{figure}
\centering
\includegraphics[width=\hsize]{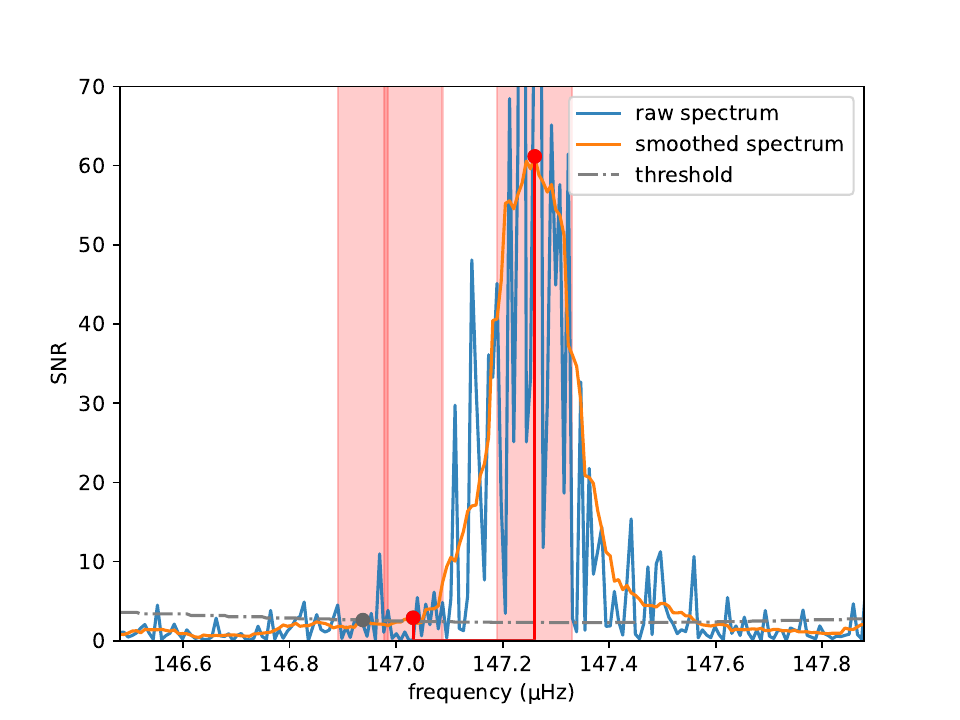}
  \caption{Same as Fig. \ref{group3b} for a p-dominated singlet in the PSD of KIC~11087371.
  }
     \label{group3c}
\end{figure}

The third step tries to eliminate cases where consecutive selected peaks belong to the same mode component. This is done in two sub-steps. First, the distance between each peak within a group is checked. Then the distance between each group is checked. If this distance is smaller than three times the expected width of the dipolar mode at its frequency, then the peak with the smallest amplitude is removed. We found that using a minimal distance of three times the mode width works best to separate split modes. 
This step is effective in about 90\% of cases, such as in Fig. \ref{group3b}. There, the widths of the peaks do not overlap, thus preserving each component of the triplet, which is the best case scenario. In Fig.~\ref{group3} and \ref{group3c}, we show typical caveats that arise in the remaining 10\% of cases.
In the first case, shown in Fig.~\ref{group3}, we see another triplet of KIC 4660930 (a $p$-dominated mode). Again, the $m=0$ and $m=-1$ components are within the same group, with an additional detection at around 156.65~$\mu$Hz. However, the width of modes overlap, which means that only the guess with the highest intensity is retained. The second case, shown in Fig. \ref{group3c}, corresponds to a high-SNR $p$-dominated mode. We see how noise in the left wing of the mode has caused an unwanted secondary detection. This detection is not removed because the red areas do not overlap. In most problematic cases, the modes involved are $p$-dominated modes because they have larger mode widths. These types of issues have to be addressed manually.

These few select cases show the limits of such a simple approach. In the global case, such cases are few and far between (around one or two modes per star) but still require additional vigilance and a thorough manual examination of the identifications after the fact. In a future study, we aim to reduce this kind of errors by refining the detection algorithm.

\section{Computing $p_{\nu_\mathrm{mag}}(0)$}
\label{matrix}

As mentioned in Sect.~\ref{modelc}, we need to compute the value of $p_{\nu_\mathrm{mag}}(0)$ from a MCMC sampling. We build a density histogram $P_h$ of the parameter $\nu_\mathrm{mag}$ with a bin size $h$ based on Scott's rule \citep{Scott79}:
\begin{equation}
    h = 3.49\frac{\sigma}{N^{1/3}}\;,
\end{equation}
where $\sigma$ is the standard deviation of $\nu_\mathrm{mag}$ and $N$ is the number of samples. This rule is generally a good choice for high count histograms, which is the case here. The value of $p(0)$ -- we dropped the index $\nu_\mathrm{mag}$ for visual clarity -- can be extrapolated from the first bin of the histogram, $P_{h}[1]$. Since the first bin is the interval $[0,h[$, $P_{h}[1]$ is related to $p$ through the relation
\begin{equation}
    P_{h}[1] \approx \frac{1}{h}\int_0^h{p(\nu_\mathrm{mag})}\hbox{d}\nu_\mathrm{mag}.
    \label{eq:hist}
\end{equation}

We expand $p$ around $0$ using a Taylor series:
\begin{equation}
    p(x) \approx p(0) + x p'(0) +\frac{x^2}{2}p''(0) + O(x^3)
    \label{eq:TaylorApp}
\end{equation}
where the script $'$ indicates the derivative. By combining Eqs.~\ref{eq:hist} and \ref{eq:TaylorApp}, and computing histograms with bin sizes $h/2$ and $2h$, we obtain the following system of equations:
\begin{gather}
    \begin{pmatrix} P_{h}[1] \\ P_{h/2}[1] \\P_{2h}[1] \end{pmatrix}
    =
    \begin{bmatrix} 1 &1&1 \\1&1/2&1/4\\1&2&4\end{bmatrix}
    \cdot
    \begin{pmatrix}p(0)\\ hp'(0)/2 \\h^2 p''(0)/6 \end{pmatrix}\;
\end{gather}
Inverting this system yields the approximated value of $p(0)$.

\end{appendix}

\end{document}